\documentclass[final,3p]{elsarticle}
\biboptions{numbers,sort&compress}

\usepackage{amssymb}
\usepackage{lineno}
\usepackage[]{siunitx}
\usepackage{multirow}
\usepackage{wasysym}
\usepackage{xcolor}
\usepackage{subcaption}
\usepackage{footnote}
\usepackage{graphicx}
\usepackage[utf8]{inputenc}
\usepackage{makecell}
\usepackage[page,toc,titletoc,title]{appendix}
\usepackage[subfigure]{tocloft}
\usepackage[section]{placeins} 

\usepackage[export]{adjustbox}

\usepackage{siunitx} 
\DeclareSIUnit\sq{\ensuremath{\Box}}                           

\usepackage[rawfloats=true]{floatrow}

\usepackage{xcolor}

\usepackage{hyperref}

\begin{document}
\begin{frontmatter}
\title{Studying AC-LGAD strip sensors from laser and testbeam measurements}

%

\author[1]{Danush Shekar\corref{cor1}}\ead{dsheka3@uic.edu}
\author[1]{Shirsendu Nanda}
\author[2]{Zhenyu Ye}
\author[2]{Ryan Heller}
\author[3]{Artur Apresyan}

\affiliation[1]{organization={University of Illinois at Chicago},
            city={Chicago},
            postcode={60607}, 
            state={IL},
            country={USA}}

\affiliation[2]{organization={Lawrence Berkeley National Laboratory},
            city={Berkeley},
            postcode={94720}, 
            state={CA},
            country={USA}}

\affiliation[3]{organization={Fermi National Accelerator Laboratory},
            addressline={PO Box 500}, 
            city={Batavia},
            postcode={60510-5011}, 
            state={IL},
            country={USA}}


\begin{abstract}
This paper presents the setup assembled to characterize and measure the spatial and timing resolutions of AC-coupled Low Gain Avalanche Diodes (AC-LGADs), using a 1060 nm laser source to deposit initial charges with a defined calibration methodology. The results were compared to those obtained with a 120 GeV proton beam. Despite the differences in the charge deposition mechanism between the laser and proton beam, the spatial and temporal resolutions were found to be compatible between the two sources after calibration. With 4D tracking detectors expected to play a vital role in upcoming collider experiments, we foresee this work as a way to evaluate the performance of semiconductor sensors that can augment testbeam measurements and accelerate R\&D efforts. Additionally, simulation studies using Silvaco TCAD and Weightfield2 were carried out to understand the various contributing factors to the total time resolution in AC-LGAD sensors, measured using the laser source.
\end{abstract}

\end{frontmatter}
\tableofcontents

\section{Introduction}\label{sec:intro}
The CMS and ATLAS experiments will undergo upgrades for the High-Luminosity Large Hadron Collider (HL-LHC), which will benefit from precise timing information with the addition of new timing layer detectors to help separate out-of-time pileup tracks~\cite{CMSetd_2019, ATLAShgtd_2020}. These detectors are made of Low-Gain Avalanche Diodes (LGADs)~\cite{Pellegrini_2014}, a class of silicon sensors with an internal gain layer responsible for generating sufficiently high electric fields to enable charge multiplication via impact ionization. This internal amplification, combined with the thinness of the sensors, are key factors enabling LGADs to achieve time resolutions on the order of 10~ps.

With high energy experiments entering regimes of increasing luminosities and number of pileup interactions, there is a growing need to integrate precise timing and fine spatial measurements into detectors. 4D trackers with fast timing and fine spatial performance have been shown to be helpful in such environments~\cite{Berry:2807541, Nakamura:2021uqg}. While LGADs are well-suited for providing precise time information, their position resolution is limited by the minimum pixel size, which is set by the need for additional implants, such as junction termination edges, around the pixel edges. These implants create dead zones that reduce the overall fill factor~\cite{Giacomini_2021}. AC-LGADs~\cite{Mandurrino_2019} are designed to overcome the limitations of traditional LGADs by replacing the segmented n+ layer with a continuous resistive n+ layer. This design, combined with a continuous gain and dielectric layer, enables a 100\% fill-factor. AC-LGADs are also characterized by their signal sharing mechanism that improves position resolution~\cite{Tornago_2021} while maintaining timing performance comparable to LGADs. Given these capabilities, the AC-LGAD technology is emerging to be a promising candidate for 4D tracking systems~\cite{Madrid:2022rqw, Dutta:2024ugh}. In particular, it is a potential candidate for the electron-ion collider~\cite{aclgadForEIC_2023}, where it would be the first demonstration of 4D tracking layers in a collider experiment.

As AC-LGAD technology is in the early stages of R\&D, ongoing studies~\cite{Dutta:2024ugh} at testbeam facilities to understand and evaluate the performance (position and time resolution) of these sensors are being conducted using incident Minimum Ionizing Particles (MIPs). In addition to test beam measurements, an in-house setup using a laser source that can evaluate AC-LGAD sensor performance would be beneficial to expedite research and development efforts. Laser setups also expand the possibility of performing additional tests, leveraging the precise control of the incident source position.

An experimental setup was constructed to perform AC-LGAD position and timing resolution measurements using an infrared laser source. Section~\ref{sec:exp_methods} describes the experimental setup along with the procedure and offline analysis conducted to obtain these measurements. Following the calibration methodology defined in Section~\ref{sec:calibration-methodology}, results obtained using this setup are compared with results from a testbeam campaign~\cite{Dutta:2024ugh} in Section~\ref{sec:results}. These experimental measurements were investigated through simulations to develop an understanding of the contributing sources to the time resolution in Section~\ref{sec:scale-factors}. Finally, concluding remarks and plans for future studies are presented in Section \ref{sec:discussion}.

\section{Experimental setup and analysis methodology}\label{sec:exp_methods}

\subsection{AC-LGAD sensors}\label{sec:sensors}
\begin{figure}[h]
    \centering
    \includegraphics[width=2in, height=1.5in]{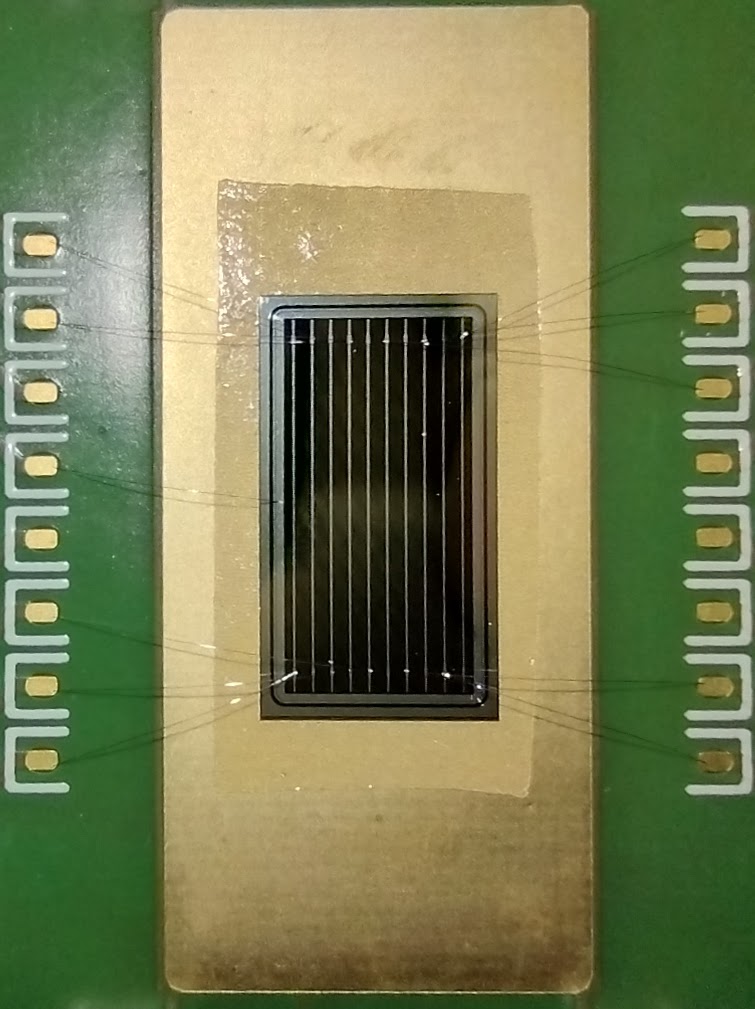} 
    \caption{An image of a HPK strip-type sensor.}
    \label{fig:hpk-sensor-img}
\end{figure}

A large variety of strip and pixel AC-LGAD sensors were fabricated by Hamamatsu Photonics K.K. (HPK) and Brookhaven National Laboratory to understand the impact of sheet resistance of the n$+$ layer, coupling capacitance, active thickness, and readout electrode geometry on sensor performance. The properties of resistive and capacitive layers present underneath the metal readout electrodes in AC-LGAD sensors influence the signal characteristics and signal sharing mechanisms. The findings were reported in~\cite{Dutta:2024ugh} along with details on the production campaign. This study will focus on a subset of strip-type AC-LGAD sensors from HPK as listed in Table \ref{tab:sensor-info-strips}. Each sensor comprises ten strips that are each 1 cm long and 50 $\mu$m wide with a pitch of 500 $\mu$m. An image of a strip-type sensor from HPK, wire-bonded to the readout board is shown in Figure \ref{fig:hpk-sensor-img}. While all ten strips were wire-bonded to the readout chains, signals from only three of them were recorded using an oscilloscope.

\begin{table}[htp]
\centering
\begin{tabular}{| c| c| c| c| c| c| c| c| c| c |}
\hline
 \multicolumn{1}{|c|}{Name} & \multicolumn{1}{c|}{\begin{tabular}[c]{@{}c@{}} Alias\\used\end{tabular}} & \multicolumn{1}{c|}{Wafer} &  \multicolumn{1}{c|}{Pitch} & \multicolumn{1}{c|}{\begin{tabular}[c]{@{}c@{}} Strip\\length \end{tabular}} & \multicolumn{1}{c|}{\begin{tabular}[c]{@{}c@{}} Metal\\width \end{tabular}} & \multicolumn{1}{c|}{\begin{tabular}[c]{@{}c@{}} Active\\thickness \end{tabular}} & \multicolumn{1}{c|}{\begin{tabular}[c]{@{}c@{}}Sheet\\resistance\end{tabular}} & \multicolumn{1}{c|}{\begin{tabular}[c]{@{}c@{}} Coupling\\Capacitance \end{tabular}} &  \begin{tabular}[c]{@{}c@{}} Optimal bias\\voltage \end{tabular} \\
& in~\cite{Dutta:2024ugh} &  & \multicolumn{1}{c|}{\si{[\um]}} & \multicolumn{1}{c|}{\si{[\mm]}} & \multicolumn{1}{c|}{\si{[\um]}} &  \multicolumn{1}{c|}{\si{[\um]}} & \multicolumn{1}{c|}{\si{[\Omega / \sq]}} & \multicolumn{1}{c|}{\si{[\pico F / \mm^2]}} &  \si{[\V]} \\
\hline
\hline
S1   & SH4    & W2    & \multirow{4}{*}{500}                                      & \multirow{4}{*}{10} & \multirow{4}{*}{50} & 50        & 1600 & 240 & 198          \\\cline{1-3}\cline{7-10}
S2 & SH2    & W4 &  &  &  & 50 & 400  & 240 & 222 \\\cline{1-3}\cline{7-10}
S3 & SH5    & W5 &  &  &  & 50 & 1600 & 600 & 205 \\\cline{1-3}\cline{7-10}
S4 & SH1    & W9 &  &  &  & 20 & 1600 & 600 & 121 \\
\hline
\end{tabular}
\caption{List of the aliases and characteristic properties of the sensors used in this study.}
\label{tab:sensor-info-strips}
\end{table}

\subsection{Setup description}
\begin{figure}[h]
    \centering
    \begin{subfigure}[b]{0.45\textwidth}
        \includegraphics[width=\textwidth]{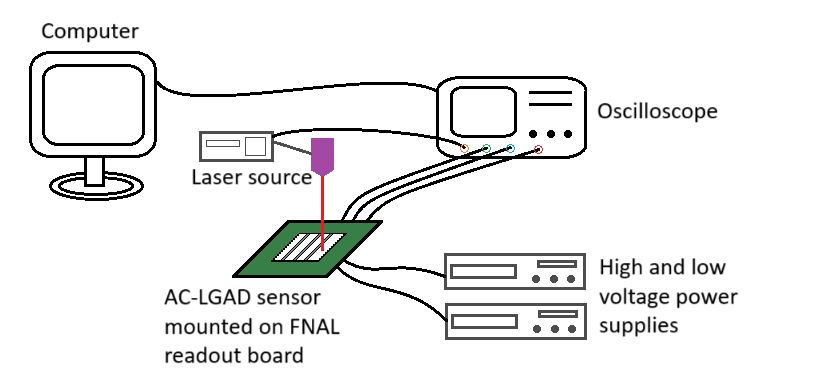}
        \caption{}
        \label{fig:setup-schematic-1}
    \end{subfigure}
    \hfill
    \begin{subfigure}{0.45\textwidth}
        \includegraphics[width=1.9in]{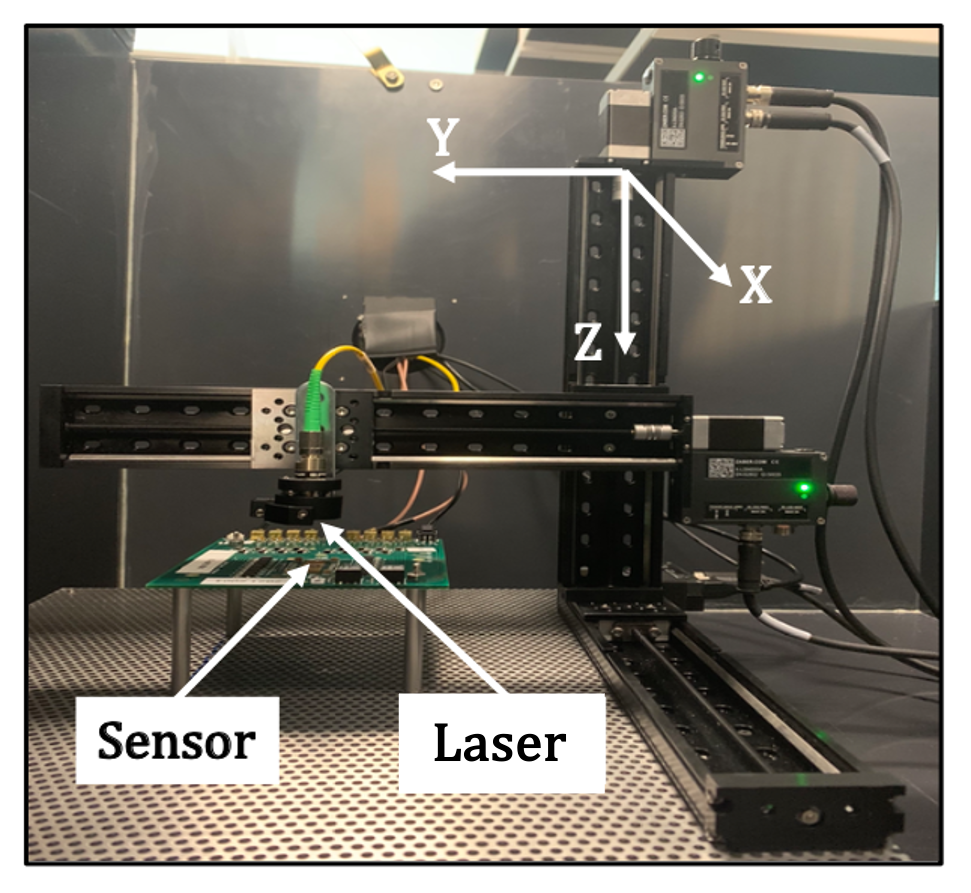}
        \caption{}
        \label{fig:setup-schematic-2}
    \end{subfigure}
    \caption{(a) A schematic representation of various components of the experimental setup and (b) an image of the laser test setup.}
    \label{fig:setup-schematic}
\end{figure}

A schematic of the constructed setup is shown in Figure~\ref{fig:setup-schematic-1}. The sensor bias voltage and power for the readout electronics were supplied by high-voltage and low-voltage power supplies, respectively. A pulsed infrared laser source from NKT Photonics (PILAS DX PIL106) with low timing jitter (\textless~3~ps) was used to generate signals in the sensor emulating a MIP response. These signals generated in the sensor strips were read out using a 16-channel board designed by Fermilab (FNAL)~\cite{Dutta:2024ugh, Heller:2021mht}. Each readout chain includes a two-stage amplifier with an equivalent transimpedance of approximately 4.3~k$\Omega$, which amplifies the signal output before it is sent to the oscilloscope.

Two oscilloscopes were used during data-taking: a Keysight MSO7104B (4 analog channels, 4~GSa/s sampling rate, 1~GHz bandwidth) and a Teledyne LeCroy Waverunner 8208HD (8 analog channels, 10~GSa/s sampling rate, 2~GHz bandwidth). Data for sensor S3 were recorded using the Keysight MSO7104B, while the Waverunner 8208HD was used for the remaining sensors, as its larger bandwidth is capable of better resolving the rising edge of the signal waveforms. The oscilloscope was triggered using the trigger-out signal from the laser source, and the scope data was stored for offline analysis on a computer running LabView~\cite{ni_labview}. The same computer was also interfaced with an $XYZ$ motorized stage (consisting of three X-LRM linear stages from Zaber) to control the laser position along the $X$, $Y$, and $Z$ axes, with a precision of $\mathcal{O}$(0.1~$\mu$m).

The Device Under Test (DUT) was placed parallel to the $XY$ plane, with $Z$ defined as the direction normal to the sensor plane, and $X$ ($Y$) perpendicular (parallel) to the strip length. Two-dimensional scans were performed along the $XY$ plane with step sizes of 50~$\mu$m in $X$ and 100~$\mu$m in $Y$. For each ($X$, $Y$) position, the DUT output was recorded for 50 laser pulses. The optimal $Z$-coordinate of the laser source was determined to be that position with the sharpest change in signal amplitude as the laser source is translated (perpendicular to the strip length direction) across the edge of the readout electrode. This ensured that the laser spot size on the DUT surface was minimized, achieving a spot size of approximately 20~$\mu$m.


\subsection{Reconstruction techniques}
The offline analysis for each sensor entailed calculating both the incident position and the laser pulse's Time Of Arrival (ToA) using reconstruction techniques outlined in this section. Subsequently, these results are compared with ones obtained using the FNAL testbeam setup that was reported in~\cite{Dutta:2024ugh}.

\subsubsection{Position reconstruction}
Precise determination of the charged particle hit position on the sensor relies on the $x$‐position reconstruction method, leveraging the signal-sharing mechanism inherent to AC-LGADs. In this approach, the two strips with the highest signal amplitudes are first identified, and the hit position is interpolated between them using the amplitude fraction
\begin{equation}
    f = \frac{a_1}{a_1 + a_2},
\end{equation}
where \(a_1\) and \(a_2\) denote the amplitudes of the leading and subleading strips, respectively. Template histograms of the distance of the hit from the leading strip center (impact parameter) as a function of \(f\) are then constructed using a tracker reference and fitted with a polynomial function \(h(f)\), providing a mapping from \(f\) to the impact parameter. For each event, the measured \(f\) is used in \(h(f)\) to determine the impact parameter relative to the leading strip. This reconstruction is applied only when at least two strips exhibit a signal above a predefined noise threshold ($\sim$ 15 mV). This method is analogous to those employed in test beam measurements and has been validated in several studies~\cite{Heller:2021mht, Madrid:2022rqw, Dutta:2024ugh}. Note that, unlike proton beams, the infrared photons do not penetrate the metalized strips, thereby limiting position reconstruction in those areas.

The expected position resolution for two-strip $x$-reconstruction was presented in a previous paper~\cite{Madrid:2022rqw} by propagating the uncertainty of the noise and signal amplitudes in the two leading strips, $a_1$ and $a_2$, for the signal sharing polynomial $h(f)$:
\begin{equation}\label{eq:expected-position-resolution}
\sigma^{\text{expected}}_x = PN\left|\frac{dh}{df}\right|\frac{\sqrt{a_1^2 + a_2^2}}{(a_1 + a_2)^2}
\end{equation}
where $P$ is the sensor pitch and $N$ is the noise amplitude. As we will see, this equation will play a crucial role in clarifying and reconciling the results of spatial resolution presented in Section~\ref{sec:results}.

The signals from the AC-LGAD sensors are read out from alternating ends of the strips, and by leveraging the finite propagation times of the signal to adjacent strips, the $y$-position can be reconstructed. This technique, introduced in~\cite{Madrid:2022rqw}, demonstrated $y$-resolutions on the order of $\mathcal{O}$(mm) for centimeter-scale sensors. While this resolution is sufficient to correct for position dependent delays in time measurement, it is not accurate enough for track reconstruction~\cite{Madrid:2022rqw}. Consequently, this paper focuses exclusively on $x$-reconstruction results.

\subsubsection{Time reconstruction}\label{sec:time-reconstruction}
The reference time, $t_{0}$, is extracted from the laser trigger-out channel using a Constant Fraction Discrimination (CFD)~\cite{Spieler_CFD} algorithm with a 50\% threshold. This is done by fitting the rising edge of the waveform with a second degree polynomial. An identical threshold is applied to determine the ToA of both the leading and sub-leading channels, which are then combined through Equation~\ref{eq:sensorToA} to yield the ``multi-channel timestamp'', fully exploiting the intrinsic charge-sharing mechanism of AC-LGADs.
\begin{equation}\label{eq:sensorToA}
    t = \frac{a_1^2t_1 + a_2^2t_2}{a_1^2 + a_2^2},
\end{equation}
where $a_{1/2}$ and $t_{1/2}$ are the amplitude and ToA of the leading/sub-leading channels, respectively. The multi-channel timestamp is used when both the channels exceed the noise threshold; otherwise, the timing is determined solely from the ToA of the leading channel. The timing observable used in this analysis is defined as $\Delta t = t - t_{0}$.

We also note that there are hit-position dependent time delays originating from the finite signal propagation velocity in the resistive layer and wirebonds only at the end of the centimeter-long strips. These delays are corrected offline by creating a reference map of correction values across the sensor's active area. The reference map requires hit-position information from either an external tracker, or the reconstructed position from the sensor itself. This work employs maps calibrated by laser-actuated stage positions. More details on this correction method have been described in previous studies~\cite{Madrid:2022rqw, Dutta:2024ugh}.

The time resolution is quantified by the difference between sensor ToA and the trigger reference. Previous MIP-source studies have established that Landau fluctuations and jitter constitute major contributions to the time resolution in sensors of LGAD technology~\cite{Dutta:2024ugh, Sadrozinski_2018, Giacomini_2021}, quantifying the jitter term in AC-LGADs will be helpful in illuminating key factors influencing time resolution. The same applies to this study involving the laser source, where the jitter is expected to contribute most to the overall total time resolution. For single-channel systems like planar sensors, LGADs, etc., the jitter is described by~\cite{Sadrozinski_2018}:
\begin{equation}\label{eq:single-channel-jitter}
    \sigma_{jitter} = \frac{N}{dV/dt}
\end{equation}    
where $N$ is the baseline noise and $\frac{dV}{dt}$ is the signal slew rate. Detailed simulation studies were also carried out to examine the validity of the above equation and further described in Section~\ref{sec:simulations}. Owing to the multi-channel environment of AC-LGAD sensors, there is a requirement to adapt the definition of jitter. We thus define ``weighted-jitter''~\cite{Dutta:2024ugh} as: 
\begin{equation}\label{eq:weighted-jitter}
    \sigma_{t, jitter} = \sqrt{\frac{a_1^4\sigma_{t_1, jitter}^2 + a_2^4\sigma_{t_2, jitter}^2}{(a_1^2 + a_2^2)^2}}
\end{equation}
where $\sigma_{t, jitter}$ is the weighted jitter and $\sigma_{t_{1(2)}, jitter}$ is the single-channel jitter value from the leading(sub-leading) channel as specified by Equation~\ref{eq:single-channel-jitter}. 

\subsection{Simulation studies}\label{sec:simulations}

\begin{figure}[h]
    \centering
    \includegraphics[width=3in]{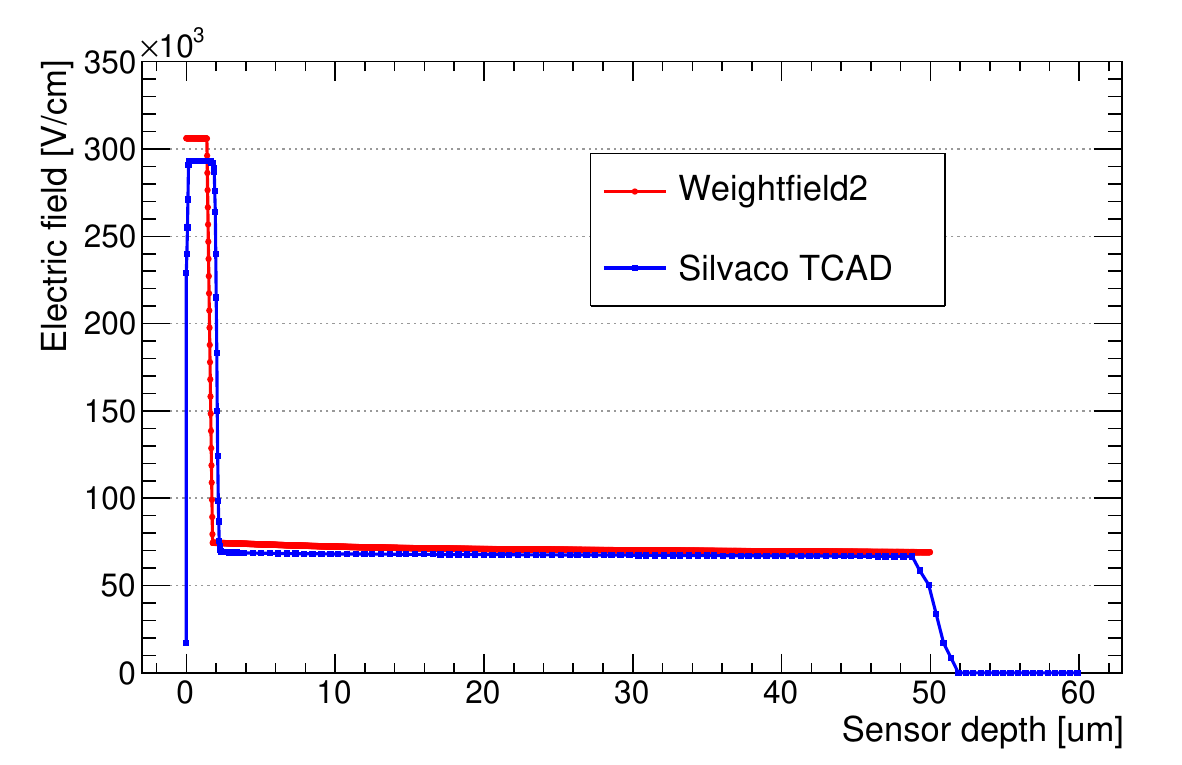} 
    \caption{A plot of the electric field profiles as a function of sensor depth simulated using WF2 and TCAD for a 50 $\mu$m thick AC-LGAD sensor. The readout electrodes and gain layer are located near sensor depth = 0 $\mu$m, while the backplane electrode of the sensor lies at high sensor depth values (50~$\mu$m in WF2 and 60~$\mu$m in TCAD).}
    \label{fig:aclgad-electric-field-profile}
\end{figure}

A simulation framework was developed using Weightfield2 (WF2)~\cite{Cartiglia_2015} and Silvaco TCAD~\cite{silvaco_tcad} that can simulate waveform shapes and estimate time resolutions. WF2 enables simulation of waveform formation—including induced signals and readout-electronics response—for AC-LGADs. Sensor parameters in WF2 were tuned to align with electric field profiles obtained from TCAD, with TCAD results based on detailed 2D process and device simulations for a 50~$\mu$m-thick sensor featuring a 10~$\mu$m low-resistivity substrate. The resulting electric field profiles are shown in Figure~\ref{fig:aclgad-electric-field-profile}, illustrating good agreement between the two profiles throughout most of the sensor bulk, and reasonably similar values in the gain layer region.

The WF2 configuration was further refined to mirror experimental conditions, including both readout-electronics and oscilloscope properties. Using the absorption coefficient of 1060~nm photons in silicon, electron-hole pairs are populated within the active volume of AC-LGAD sensors. WF2 was then used to simulate the subsequent signal propagation through the sensor and the readout chain. 

The following methodology was applied at each $x$-position as AC-LGAD signal characteristics vary with position:
\begin{enumerate}
\item Using WF2, multiple signal waveforms representing the output of the leading and subleading channels were simulated. From these, only two waveforms with risetime values matching the experimentally measured averages are saved. As amplitude and noise values are fixed from experimental inputs, risetime is chosen as the discriminating variable to select waveforms used in subsequent calculations. 
\item To eliminate event-to-event statistical fluctuations of deposited charges, 5000 copies of the two selected waveforms were created per $x$-position, so that the only source to the time resolution is noise-injected jitter.
\item Noise is injected by adding Gaussian-distributed random values to each timestamp, with zero mean and standard deviation determined by the simulated signal amplitude divided by the experimental signal-to-noise ratio.
\item Simulated waveforms were analyzed with identical procedures applied to experimental data. The measured signal amplitudes were used directly as inputs to simulation, ensuring accurate construction of the multi-channel timestamp and weighted jitter. Furthermore, the experimentally observed single- and two-channel efficiencies were applied to weight their relative contributions from single- or two-channel events~\cite{Dutta:2024ugh}, enabling the extraction of both the total time resolution and weighted jitter.
\end{enumerate}

\begin{figure}[htp]
    \centering
    \begin{subfigure}{0.33\textwidth}
        \includegraphics[height=1.6in]{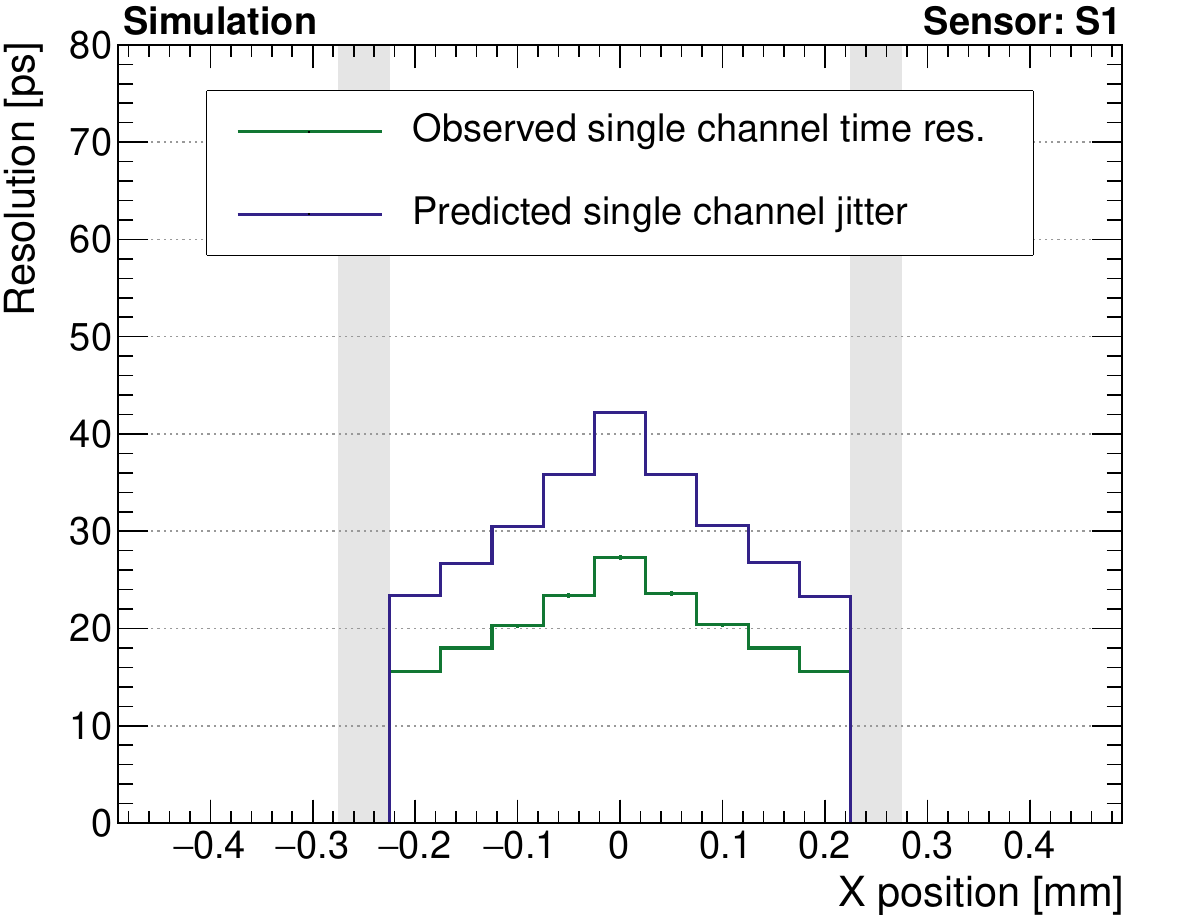}
        \caption{}
        \label{fig:simulated-tr-W2}
    \end{subfigure}
\hfill
    \begin{subfigure}{0.33\textwidth}
        \includegraphics[height=1.6in]{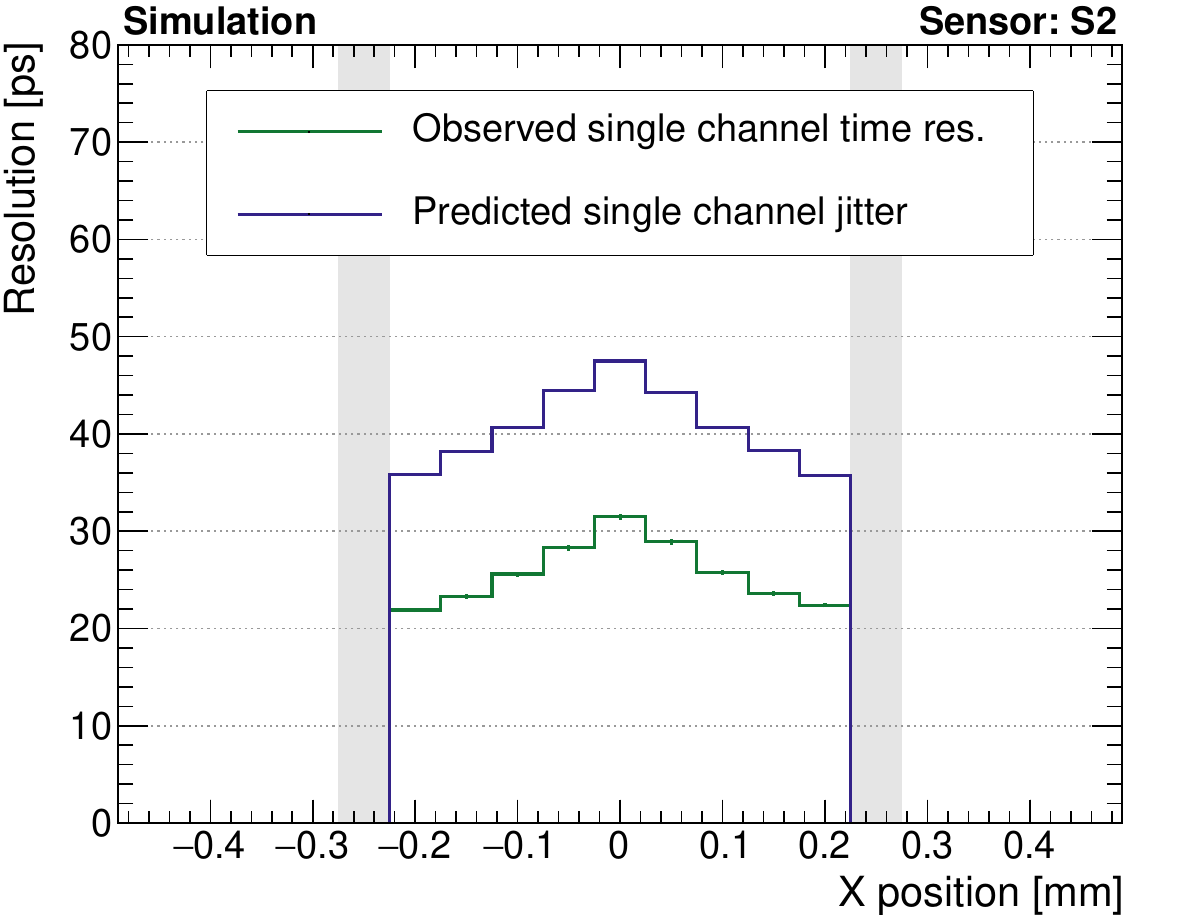}
        \caption{}
        \label{fig:simulated-tr-W4}
    \end{subfigure}
\hfill
    \begin{subfigure}{0.32\textwidth}
        \includegraphics[height=1.6in]{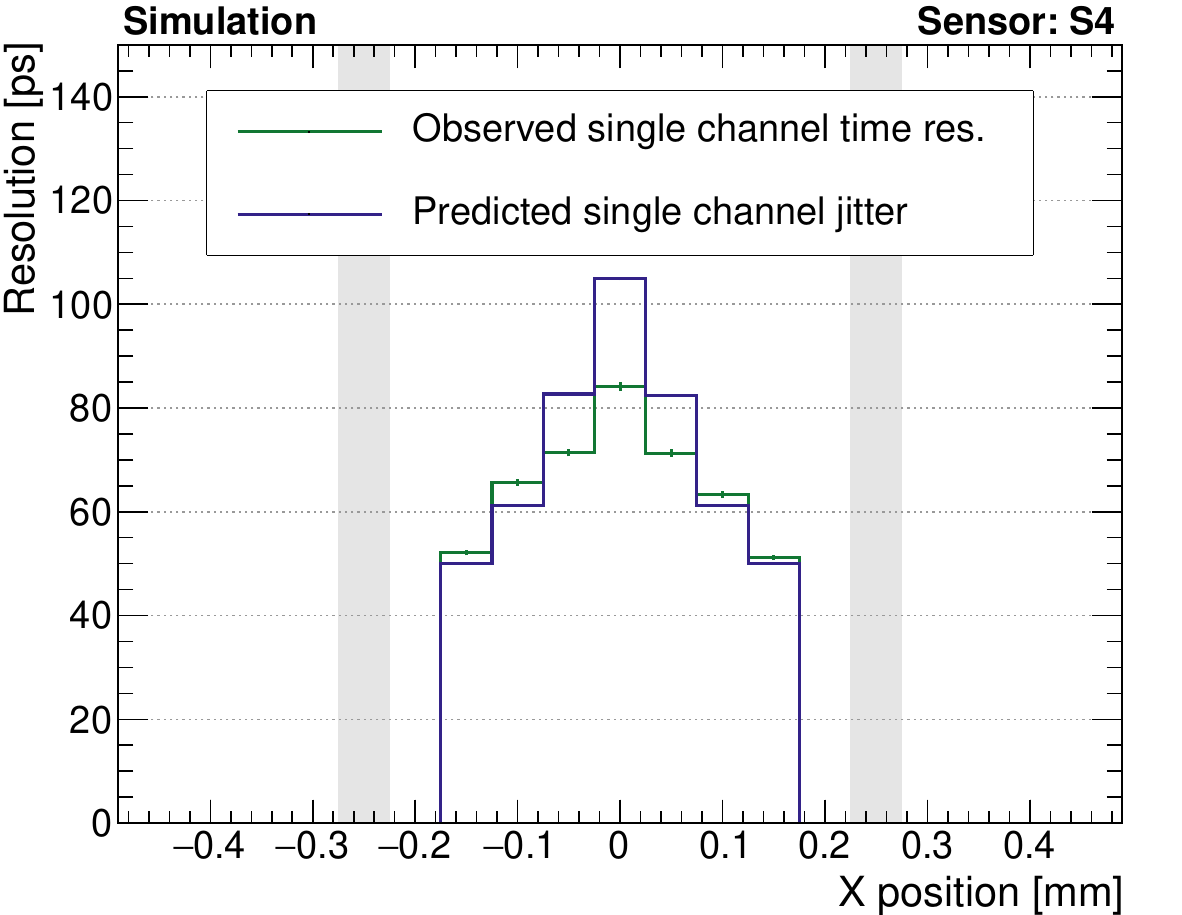}
        \caption{}
        \label{fig:simulated-tr-W9}
    \end{subfigure}
    \caption{The observed time resolution and predicted  jitter from the leading channel plotted as a function of $x$ position for sensors S1, S2, and S4, based on amplitude, noise, and efficiency values observed in the measurements.}
    \label{fig:simulated-tr-all}
\end{figure}

We compare the predicted single-channel jitter from Equation~\ref{eq:single-channel-jitter} and the observed single-channel time resolution extracted from ToA values of the waveforms, as illustrated in Figure~\ref{fig:simulated-tr-all}. We find that the predicted jitter is greater than observed time resolution, and propose a scale factor to match the observed resolution. We thus define the scale factors (SF) as the ratios of observed time resolution to predicted jitter, at each $x$ position.

\begin{figure}[htp]
    \centering
    \includegraphics[width=2.3in]{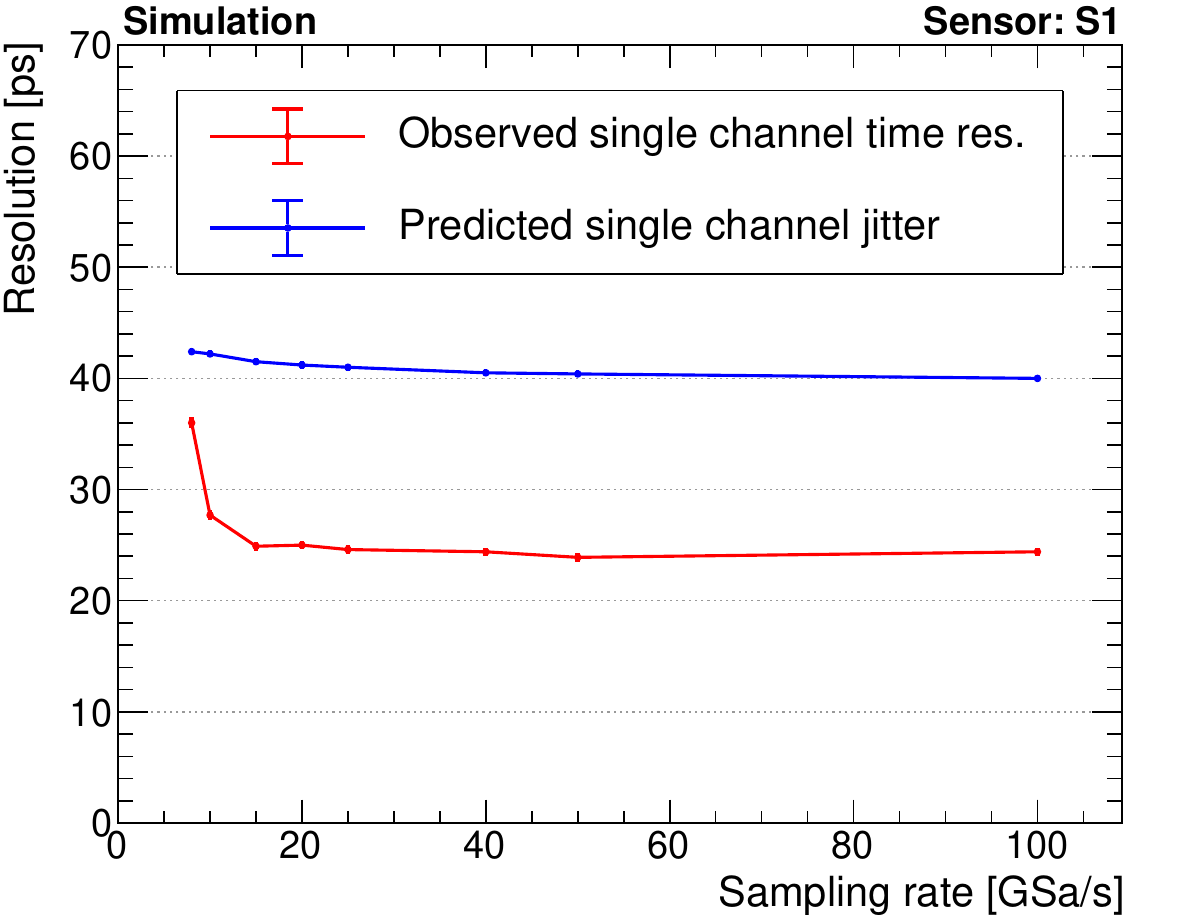}
    \caption{The observed time resolution and predicted jitter at the mid-gap position as a function of the oscilloscope sampling rate, for a 50~$\mu$m thick sensor.}
    \label{fig:sampling-rate-study}
\end{figure}

As these simulations by construction exclude contributions beyond jitter, these results are indications to inaccuracies in the single-channel jitter definition (Equation~\ref{eq:single-channel-jitter}). It should be emphasized that the contribution of jitter to the time resolution is not universal, but depends sensitively on multiple factors such as the CFD threshold and oscilloscope sampling rate~\cite{Sadrozinski_2018, Pena_2019, samplingRateOnJitter} - among others. This is highlighted in Figure~\ref{fig:sampling-rate-study}, where simulated jitter and observed time resolution are plotted against sampling rate, showing that increased sampling yields more datapoints for fitting the rising edge of the signal waveform, and thus alters resulting timing measurements. The plot was obtained by extending the simulation framework (originally designed to replicate the 10 GSa/s sampling rate of the oscilloscope used in the experiment) to evaluate these quantities at multiple sampling rates.

\section{Signal properties and calibration methodology}\label{sec:calibration-methodology}

\begin{figure}[h!]
    \centering
    \includegraphics[width=2.5in]{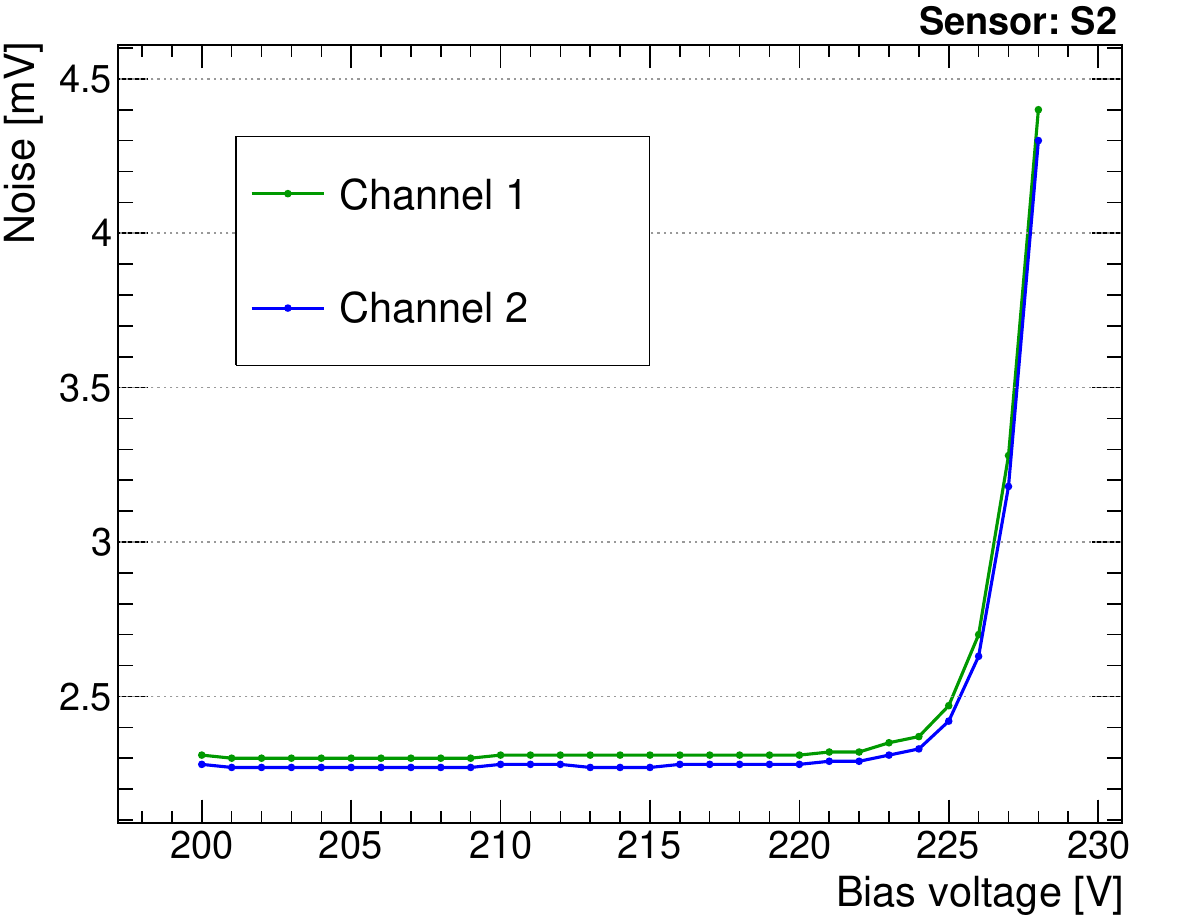}
    \caption{Noise from two channels as a function of bias voltage for Sensor S2, with 222 V identified as the operating bias.}
    \label{fig:noise-vs-bv-S2}
\end{figure}

Following the description of the experimental setup, offline analysis methodology, and simulation framework provided in the previous section, this section details the calibration methodology and some key results. All sensors were operated at a bias voltage of 1 V below breakdown and at room temperature without external cooling. The breakdown voltage for each sensor was determined by monitoring the noise level across all readout channels as a function of bias voltage. It is defined as the bias point beyond which a systematic increase in noise is observed across all channels. An example of this procedure, as performed for Sensor S2, is shown in Figure~\ref{fig:noise-vs-bv-S2}. The test beam campaign, whose results we compare with~\cite{Dutta:2024ugh}, also employed a similar procedure to determine the operating bias voltage, albeit with sensors operated at a uniform constant temperature of 20$^\circ$C.

\begin{figure}[h!]
    \centering
    \hfill
    \begin{subfigure}[b]{0.4\textwidth}
        \includegraphics[width=0.9\textwidth]{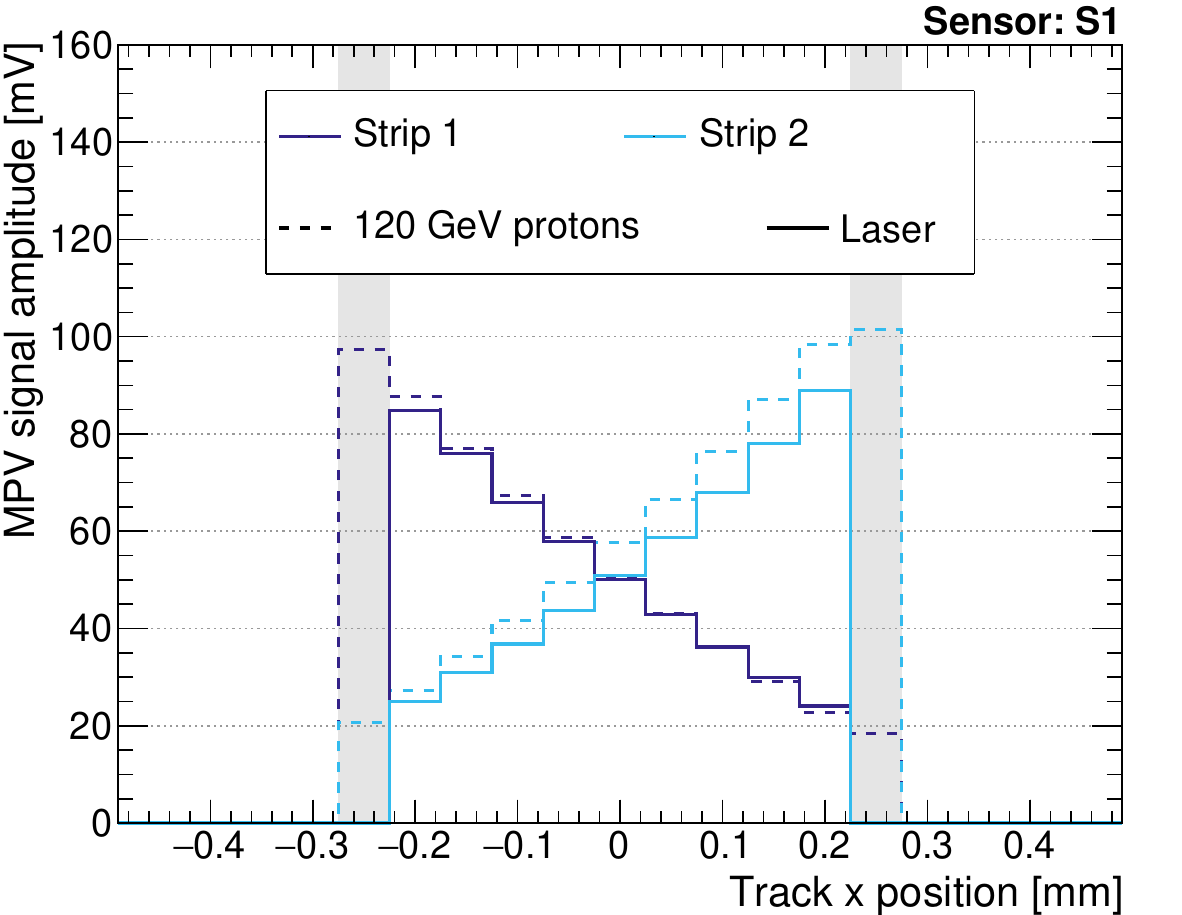}
        \caption{}
        \label{fig:amplitude-vs-x}
    \end{subfigure}
    \hfill
    \begin{subfigure}{0.4\textwidth}
        \includegraphics[width=0.9\textwidth]{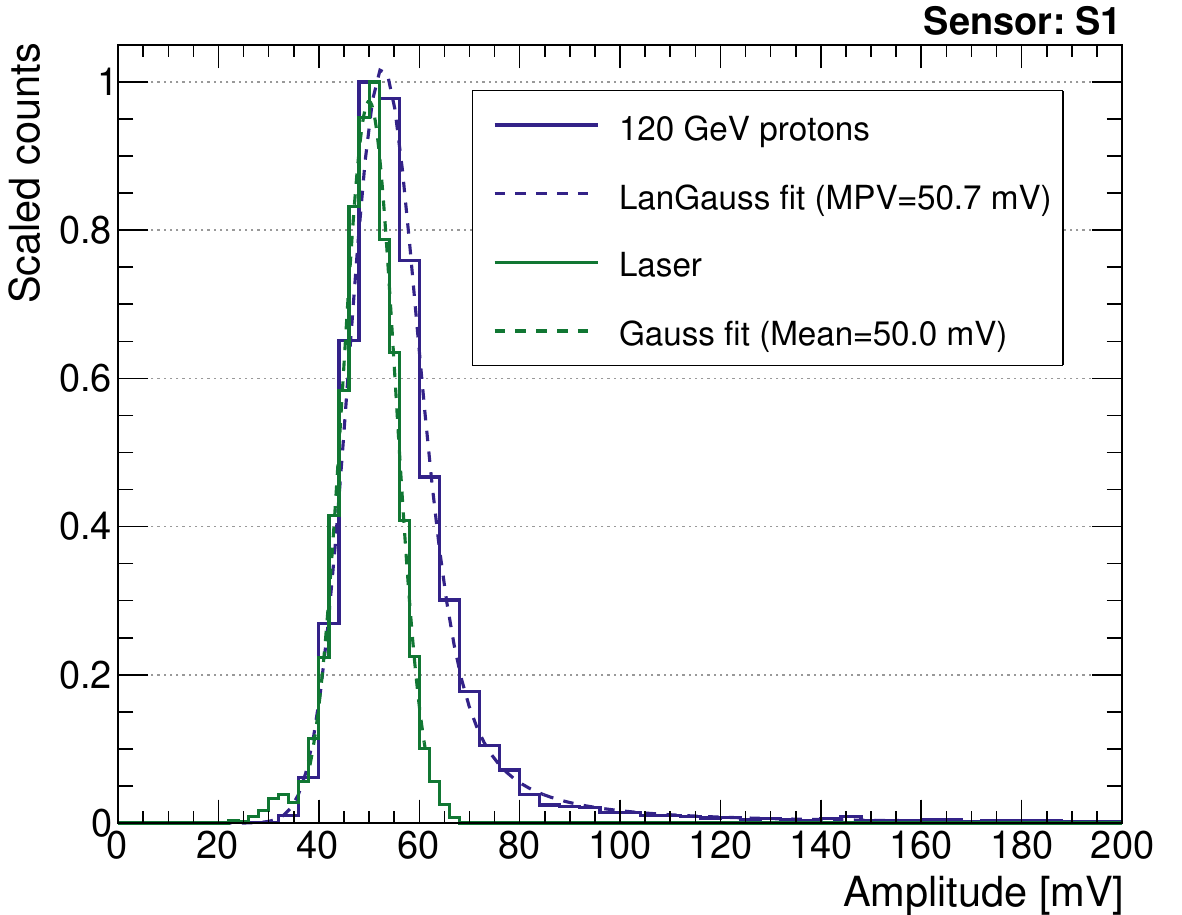}
        \caption{}
        \label{fig:amp-distrib}
    \end{subfigure}
    \caption{Comparison of (a) amplitude profile as a function of $x$-position and (b) distribution of amplitude values at mid-gap position, between MIP and laser sources.}
    \label{fig:laser-tuning}
\end{figure}

With the operational bias voltage established through the breakdown characterization described above, the next step in the calibration procedure focuses on tuning the laser intensity to reproduce the sensor response expected from MIPs. With the operational bias voltage applied to the DUT, the laser intensity is adjusted such that the mid-gap signal amplitude obtained from the laser source matches that measured from MIPs. Here, the mid-gap amplitude is defined as the signal amplitude from either of two adjacent strips when the incident particle is positioned between them (the signal is shared equally between neighboring strips). On successful tuning, the Most Probable Value (MPV) of the signal amplitude as a function of the reference tracker position ($x$) agrees across both the laser and MIP sources (as shown in Figure \ref{fig:amplitude-vs-x} for the leading channel amplitude). At individual $x$-positions (or reference tracker position, $x$), amplitude distributions for laser and MIP show distinctive shapes: Gaussian for the laser and Gaussian-convoluted Landau distribution for the MIP case (Figure~\ref{fig:amp-distrib}), consistent with their respective charge deposition mechanisms.

\begin{figure}[h!]
\centering
    \hfill
    \begin{subfigure}[b]{0.4\textwidth}
        \includegraphics[width=0.9\textwidth]{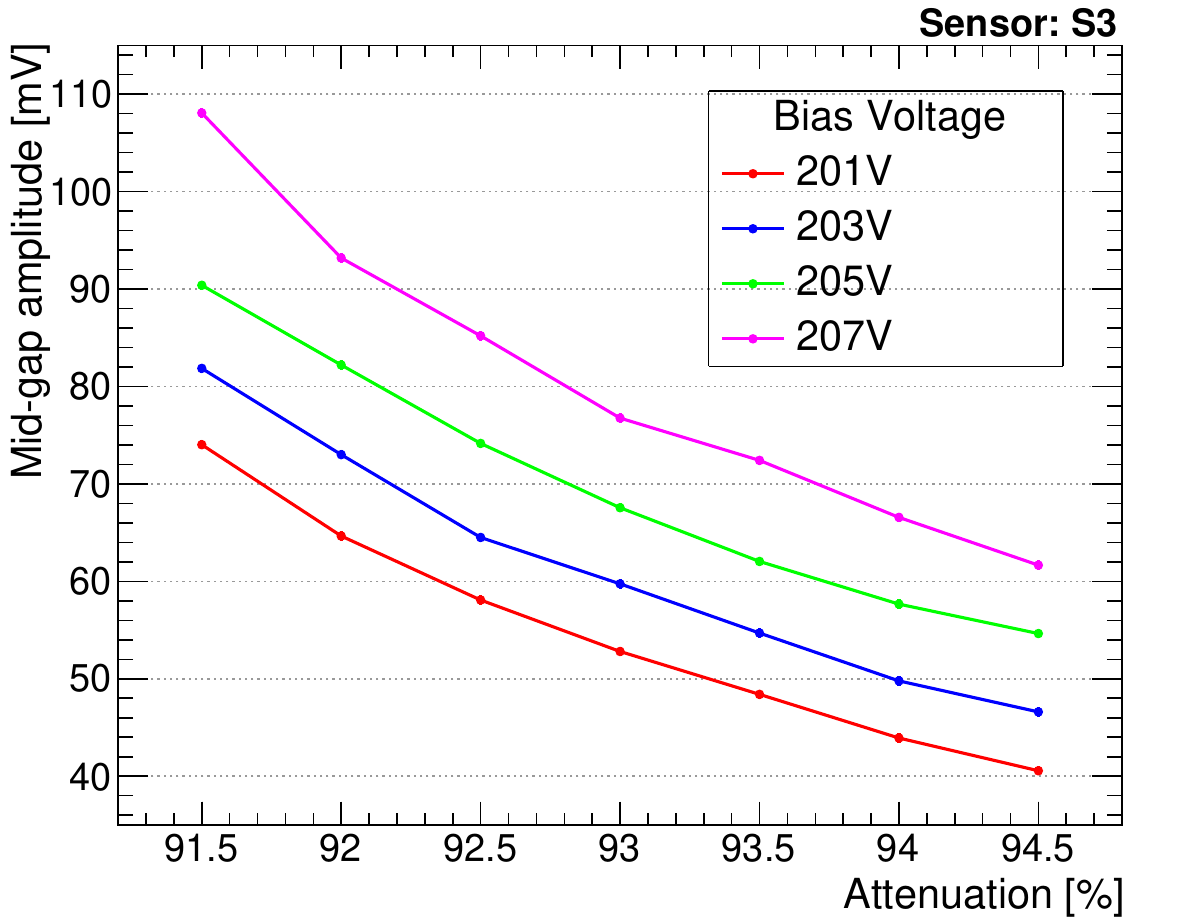}
        \caption{}
        \label{fig:midgap-amp-vs-attn}
    \end{subfigure}
    \hfill
    \begin{subfigure}{0.4\textwidth}
        \includegraphics[width=0.9\textwidth]{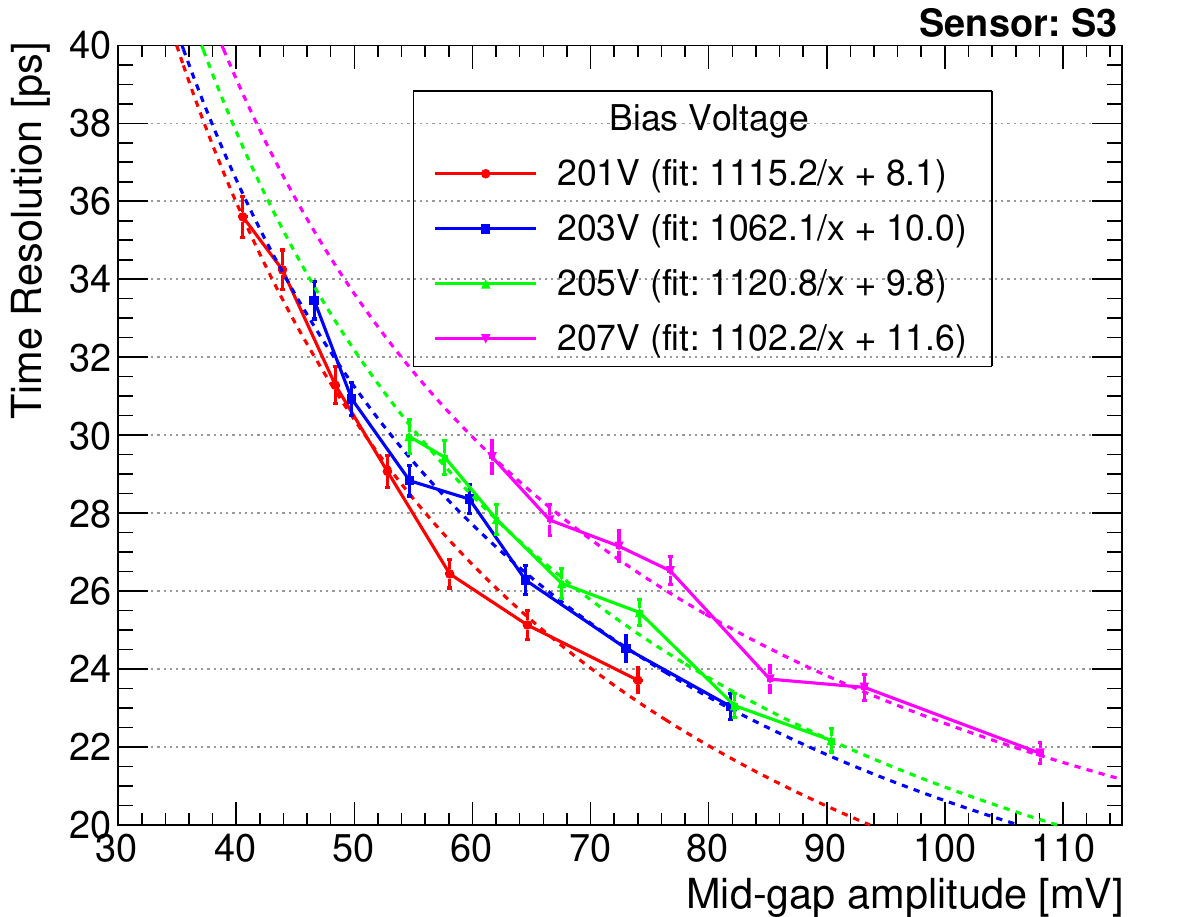}
        \caption{}
        \label{fig:midgap-amp-vs-time-res}
    \end{subfigure}
    \caption{Results exploring the phase-space of laser intensity, amplitude, and time resolution, obtained from sensor S3 using the laser source. (a) Plot of mid-gap amplitudes as a function of attenuation factor of laser intensity. (b) Plot of total time resolution at the mid-gap position as a function of the mid-gap amplitude. Multiple mid-gap amplitudes are attained by varying the laser intensity and bias voltage.} 
    \label{fig:calibration-methodology-plots}
\end{figure}

The mid-gap amplitude itself is observed to depend on both the applied bias voltage and the laser intensity, as shown in Figure~\ref{fig:midgap-amp-vs-attn}, where the $X$‑axis represents the laser intensity attenuation factor. Building on this, Figure~\ref{fig:midgap-amp-vs-time-res} further reveals that, for a given bias voltage, the time resolution follows a $\frac{a}{x}~+~b$ dependence on the mid-gap amplitude ($x$), which in turn is controlled through the laser intensity. An observed increase in noise with bias voltage is a contributing factor to the increase in time resolution with bias voltage at the same mid-gap amplitude. The presence of a non-zero $b$ term across all fits suggests a component in the time resolution distinct from jitter.

\begin{figure}[htp]
    \centering
    \begin{subfigure}{0.33\textwidth}
        \includegraphics[height=1.6in]{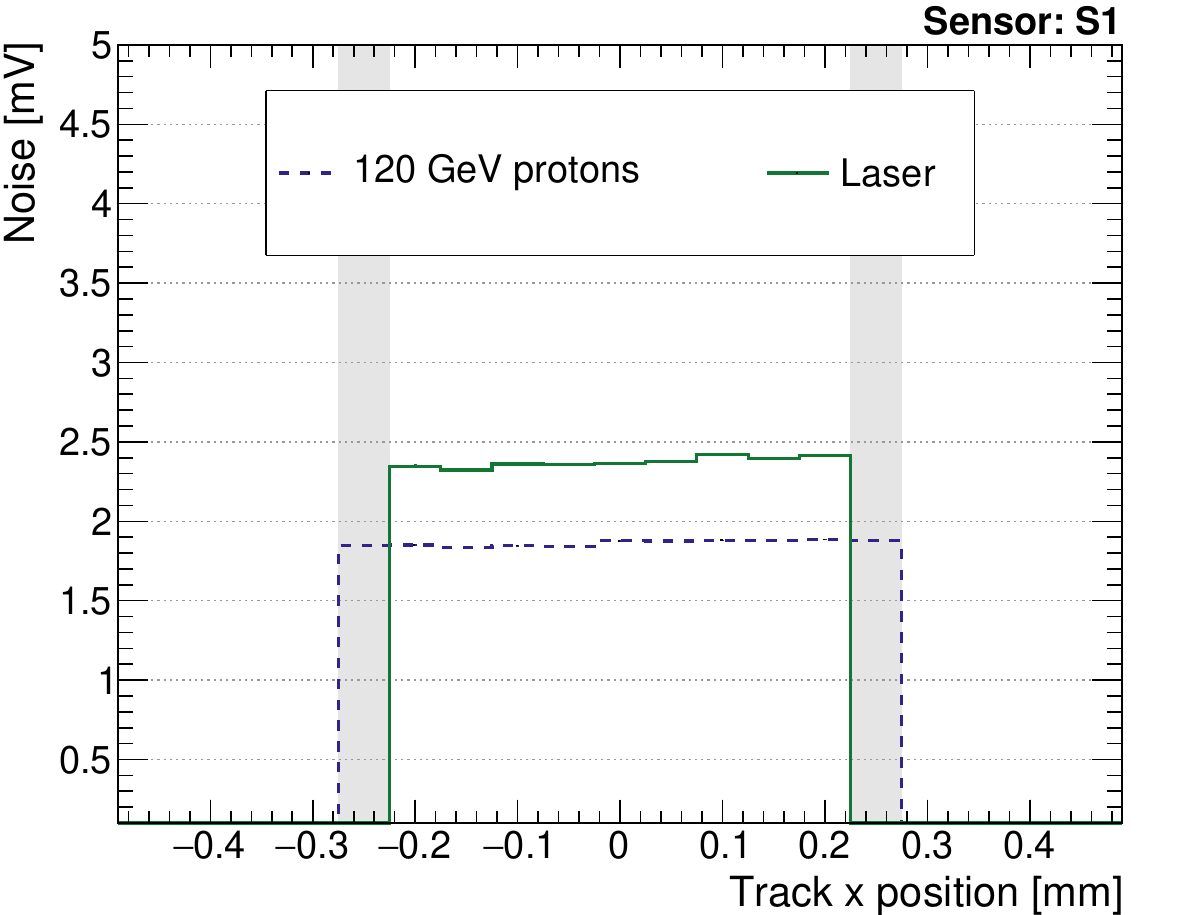}
        \caption{}
        \label{fig:noise-comparison-W2}
    \end{subfigure}
\hfill
    \begin{subfigure}{0.33\textwidth}
        \includegraphics[height=1.6in]{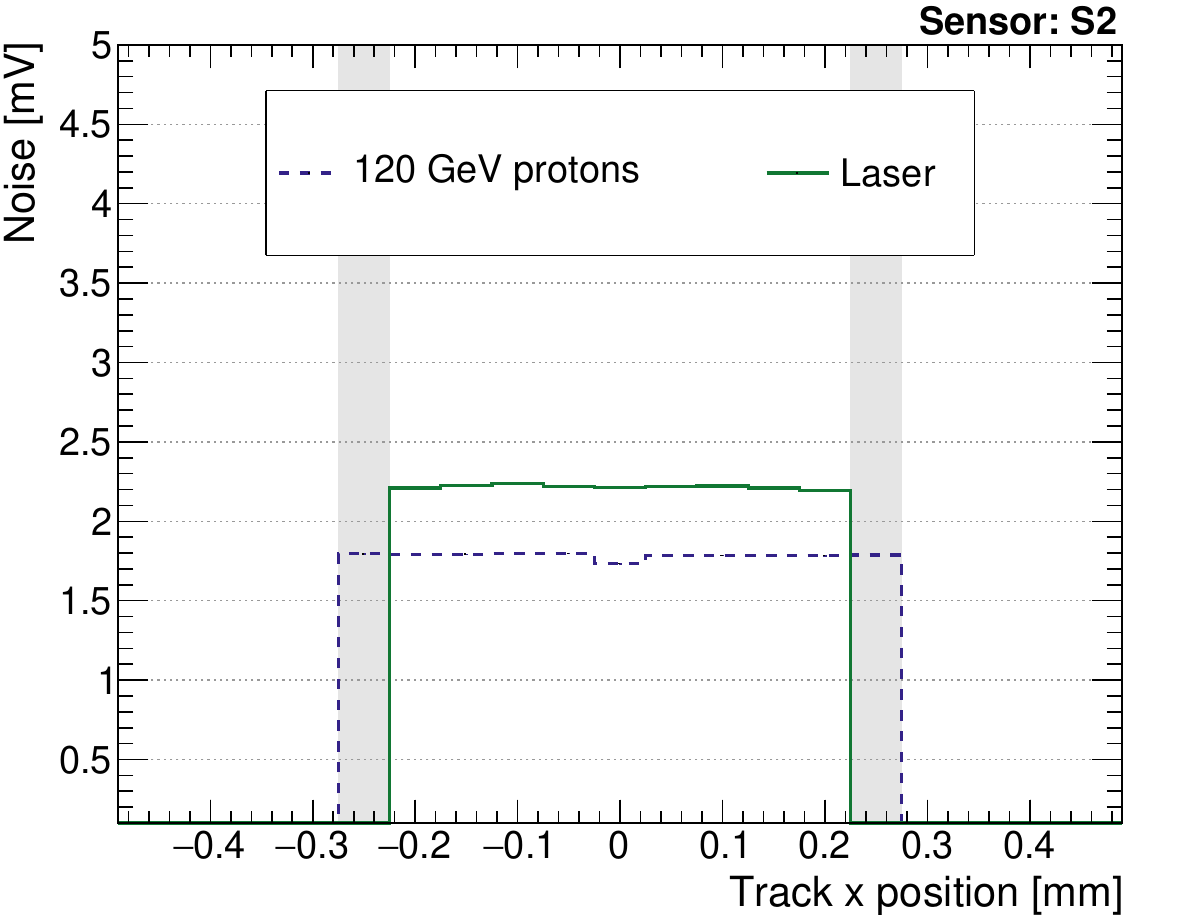}
        \caption{}
        \label{fig:noise-comparison-W4}
    \end{subfigure}
\hfill
    \begin{subfigure}{0.32\textwidth}
        \includegraphics[height=1.6in]{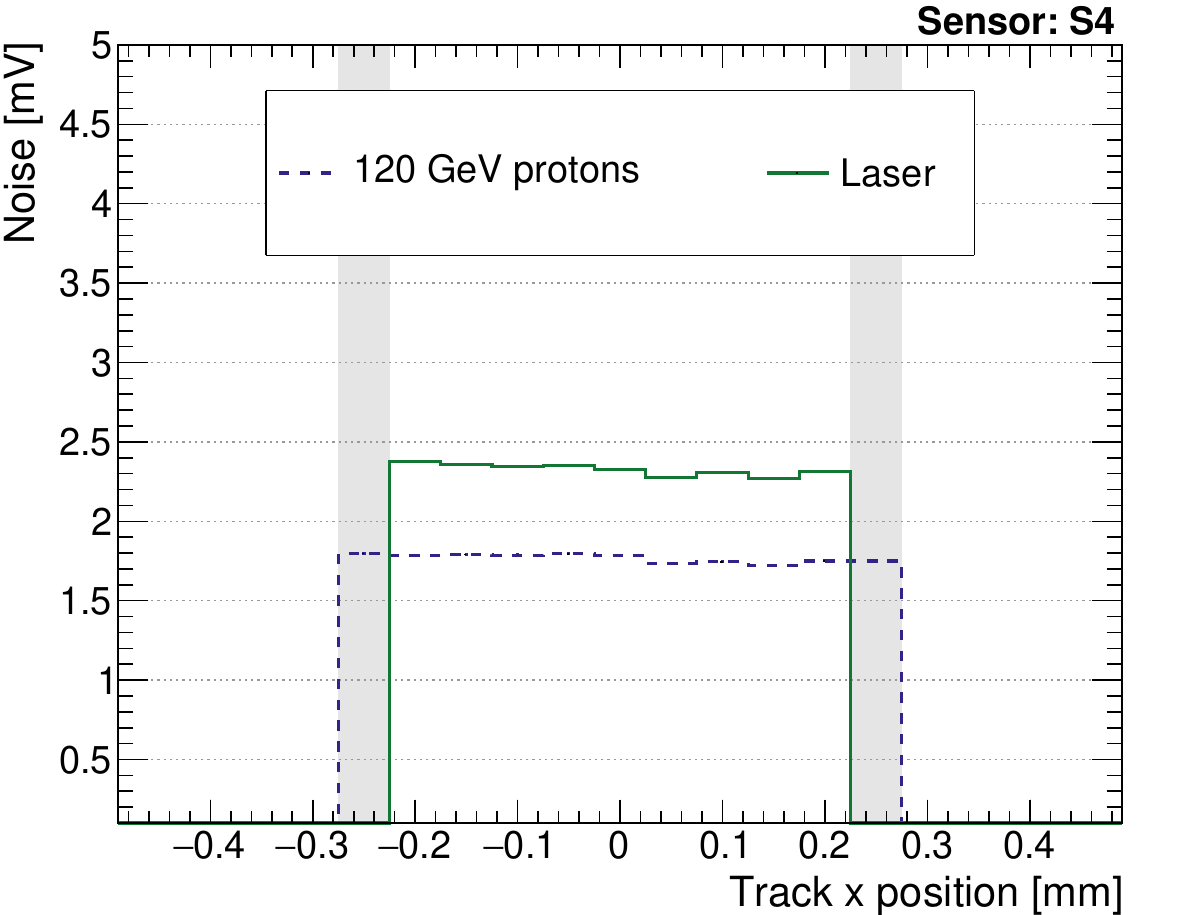}
        \caption{}
        \label{fig:noise-comparison-W9}
    \end{subfigure}
    \caption{Noise as a function of $x$ position for sensors S1, S2, and S4.}
    \label{fig:noise-comparisons-all}
\end{figure}

With the bias voltage established in accordance with the breakdown scans and the laser intensity tuned to reproduce the sensor response to MIPs, discrepancies were observed in the preliminary comparisons of the quantities of interest - spatial and temporal resolutions - between the laser and MIP sources. The source of these discrepancies was traced to differences in noise between the two setups. Figure \ref{fig:noise-comparisons-all} compares the distribution of noise as a function of track $x$ position for the laser and MIP source/setup. The differences in noise likely stem from factors like sensor temperature and the lack of optimization in the experimental setups - specifically, cable lengths, grounding of instruments, and metal components that may act as antennas. As detailed in Section~\ref{sec:results}, correcting for these noise differences restores agreement in the results of the quantities of interest between the two sources.

Notably, once the laser is calibrated using a sensor of a given thickness, the same laser intensity could be applied to another sensor of the same thickness to achieve comparable results across the quantities of interest with the MIP source. This suggests that the intensity of the laser source remained stable throughout the data-taking period for both sensors of the same thickness, which spanned approximately one day in total.

\section{AC-LGAD performance results}\label{sec:results}

Performance results from AC-LGAD sensors S1, S2, and S4, using the laser setup and the testing methodology described in the previous sections are presented here. Sensors S1 and S2 are geometrically identical and differ in the sheet resistance of the n$+$ layer. Calibration was done using amplitudes from MIP source measurements of a 50 $\mu$m thick sensor, S1, and results for S2 were obtained subsequently using this calibrated laser intensity. Furthermore, results from S4, a 20 $\mu$m thick sensor, are also shown.

\begin{figure}[htp]
    \centering
    \begin{subfigure}[b]{0.33\textwidth}
        \includegraphics[height=1.6in]{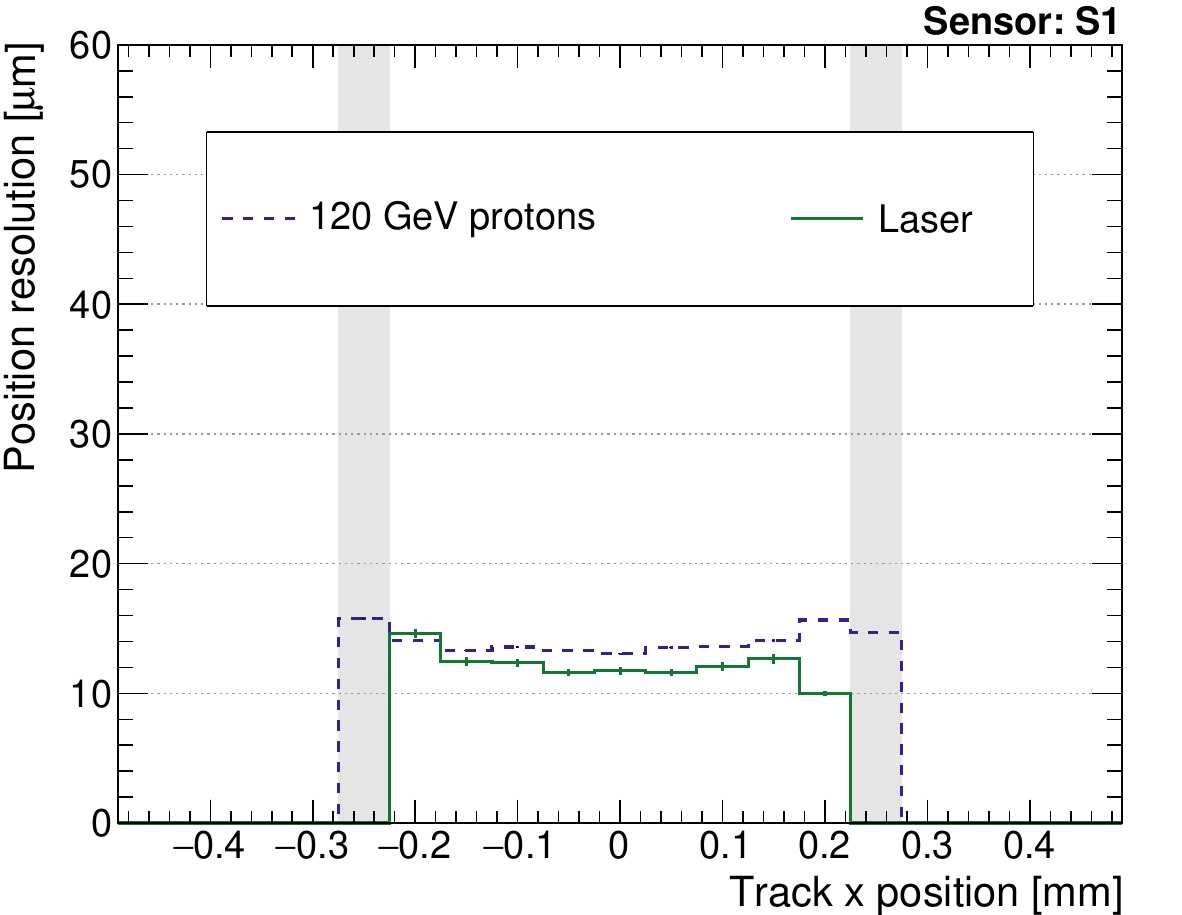}
        \caption{}
        \label{fig:pr-comparison-W2}
    \end{subfigure}
    \hfill
    \begin{subfigure}[b]{0.33\textwidth}
        \includegraphics[height=1.6in]{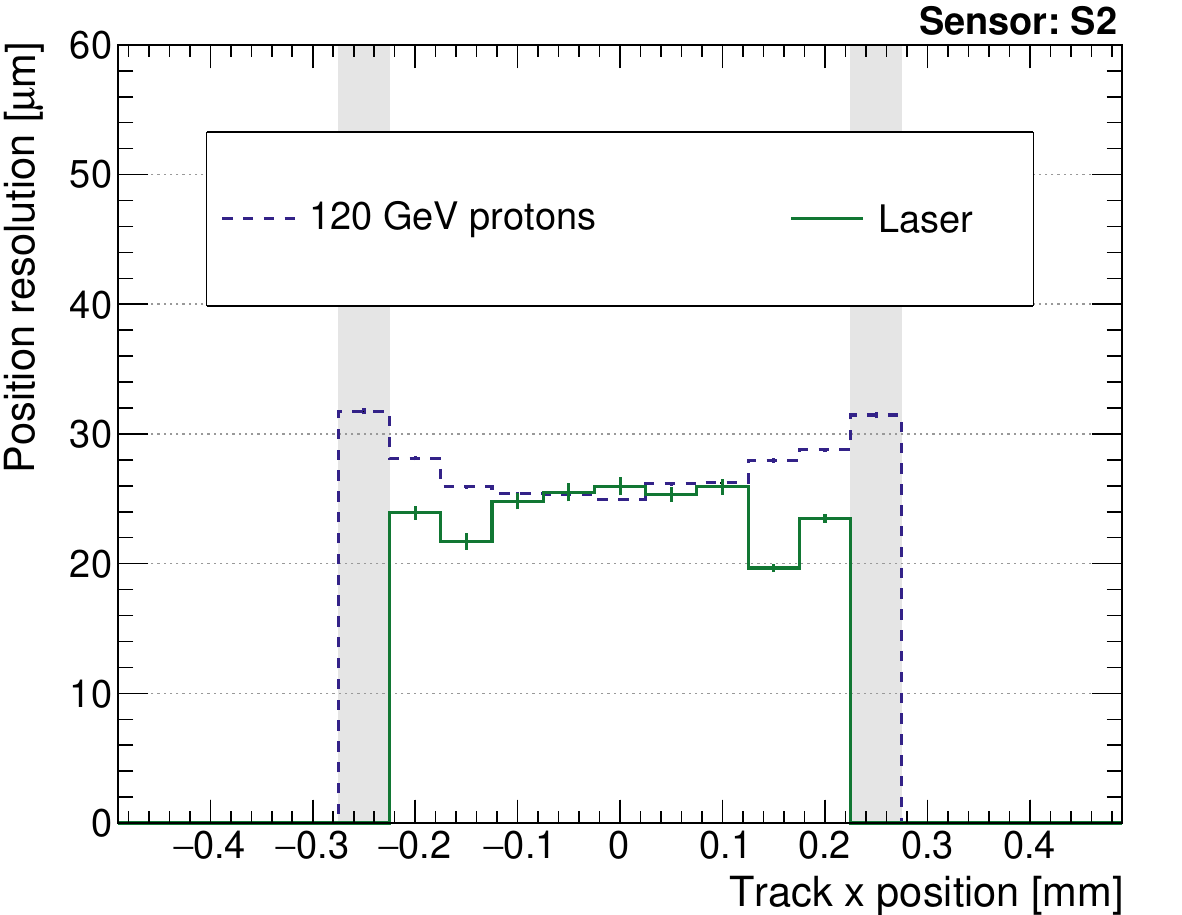}
        \caption{}
        \label{fig:pr-comparison-W4}
    \end{subfigure}
    \hfill
    \begin{subfigure}[b]{0.32\textwidth}
        \includegraphics[height=1.6in]{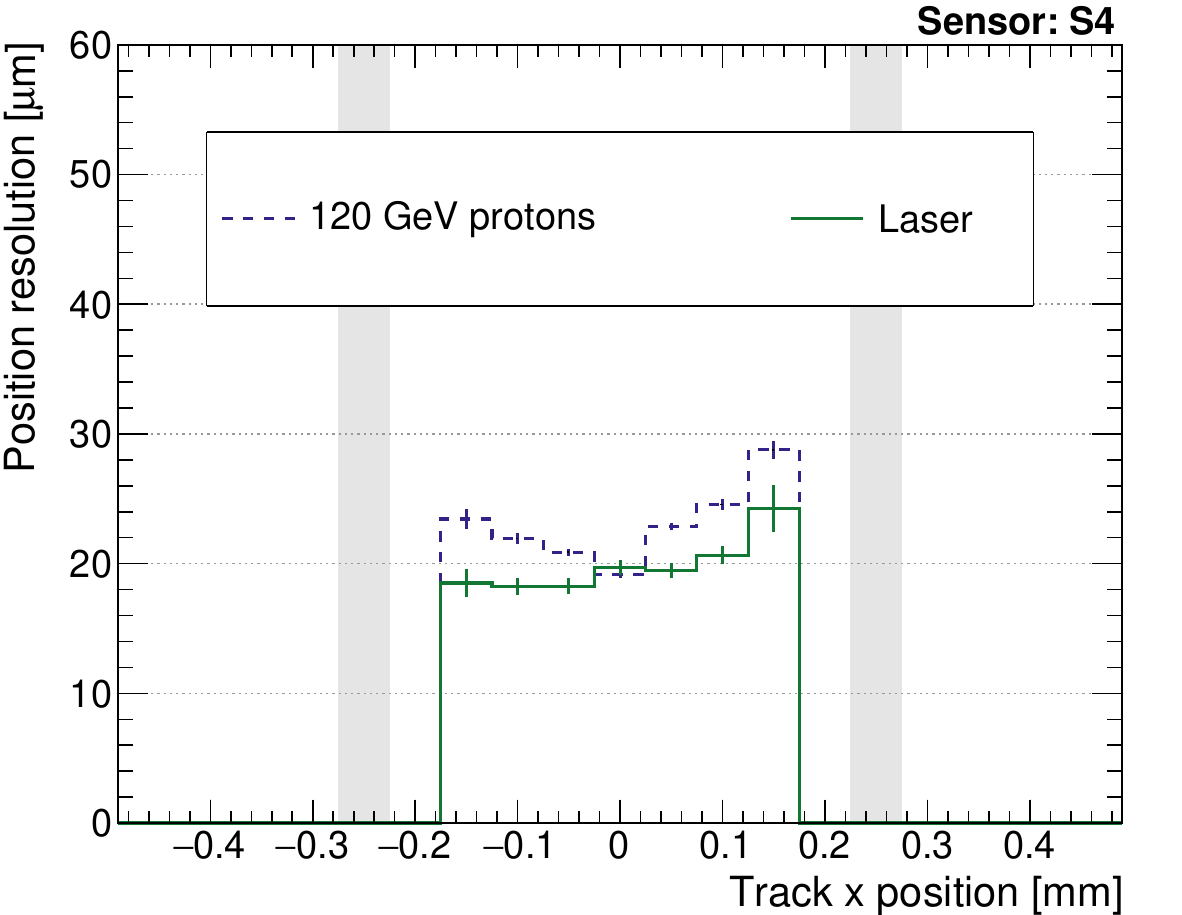}
        \caption{}
        \label{fig:pr-comparison-W9}
    \end{subfigure}
    \caption{Position resolution as a function of $x$ position for sensors S1, S2, and S4. The position resolutions from the laser source have been normalized to account for the noise differences between the two setups. Results from the 120 GeV proton beam have been reproduced with data from~\cite{Dutta:2024ugh}.}
    \label{fig:pr-comparisons-all}
\end{figure}

Figure~\ref{fig:pr-comparisons-all} contains laser and MIP source results of position resolutions (as a function of $x$ position) from all three sensors. Baseline noise differences between the setups (see Figure~\ref{fig:noise-comparisons-all}) necessitated normalizing the laser source results shown in Figure~\ref{fig:pr-comparisons-all} using Equation~\ref{eq:expected-position-resolution} (i.e. normalization factor = $\frac{N_{\text{120 GeV protons}}}{N_{\text{Laser}}}$, where $N_{\text{120 GeV protons}}$ and $N_{\text{Laser}}$ are the noise levels observed in the MIP and laser source setups, respectively). We subsequently observe the position resolutions obtained using both sources to be comparable across all three sensors as shown in Figure~\ref{fig:pr-comparisons-all}.

\begin{figure}[htp]
    \centering
    \begin{subfigure}{0.33\textwidth}
        \includegraphics[height=1.6in]{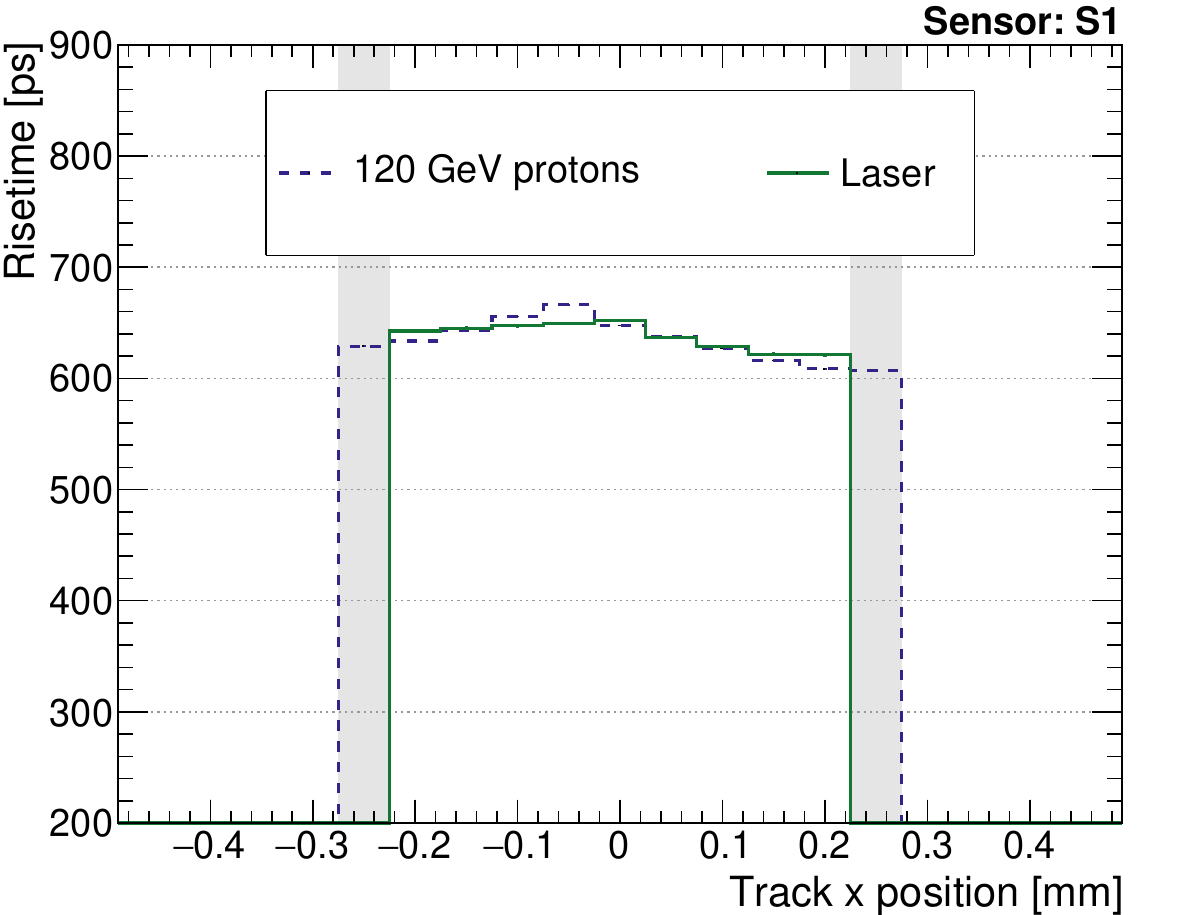}
        \caption{}
        \label{fig:risetime-comparison-W2}
    \end{subfigure}
\hfill
    \begin{subfigure}{0.33\textwidth}
        \includegraphics[height=1.6in]{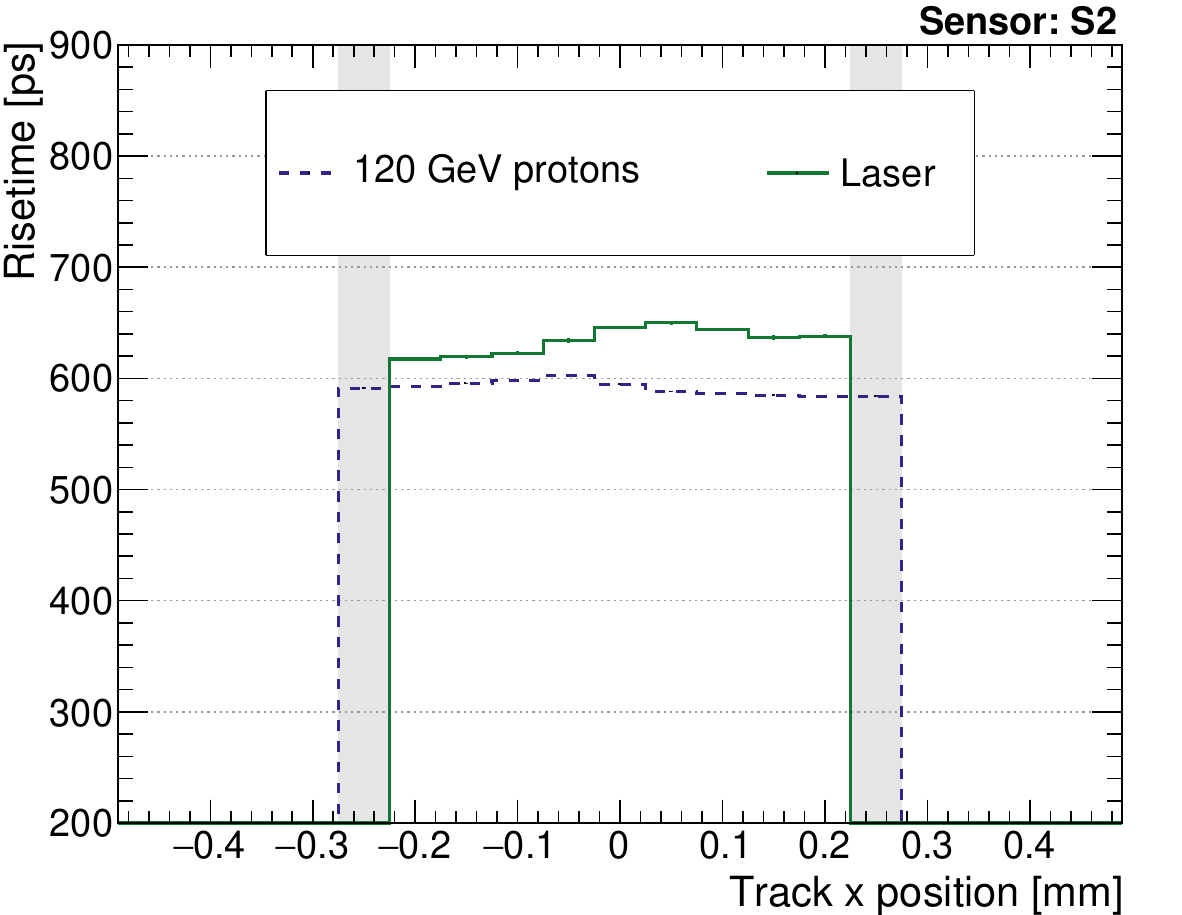}
        \caption{}
        \label{fig:risetime-comparison-W4}
    \end{subfigure}
\hfill
    \begin{subfigure}{0.32\textwidth}
        \includegraphics[height=1.6in]{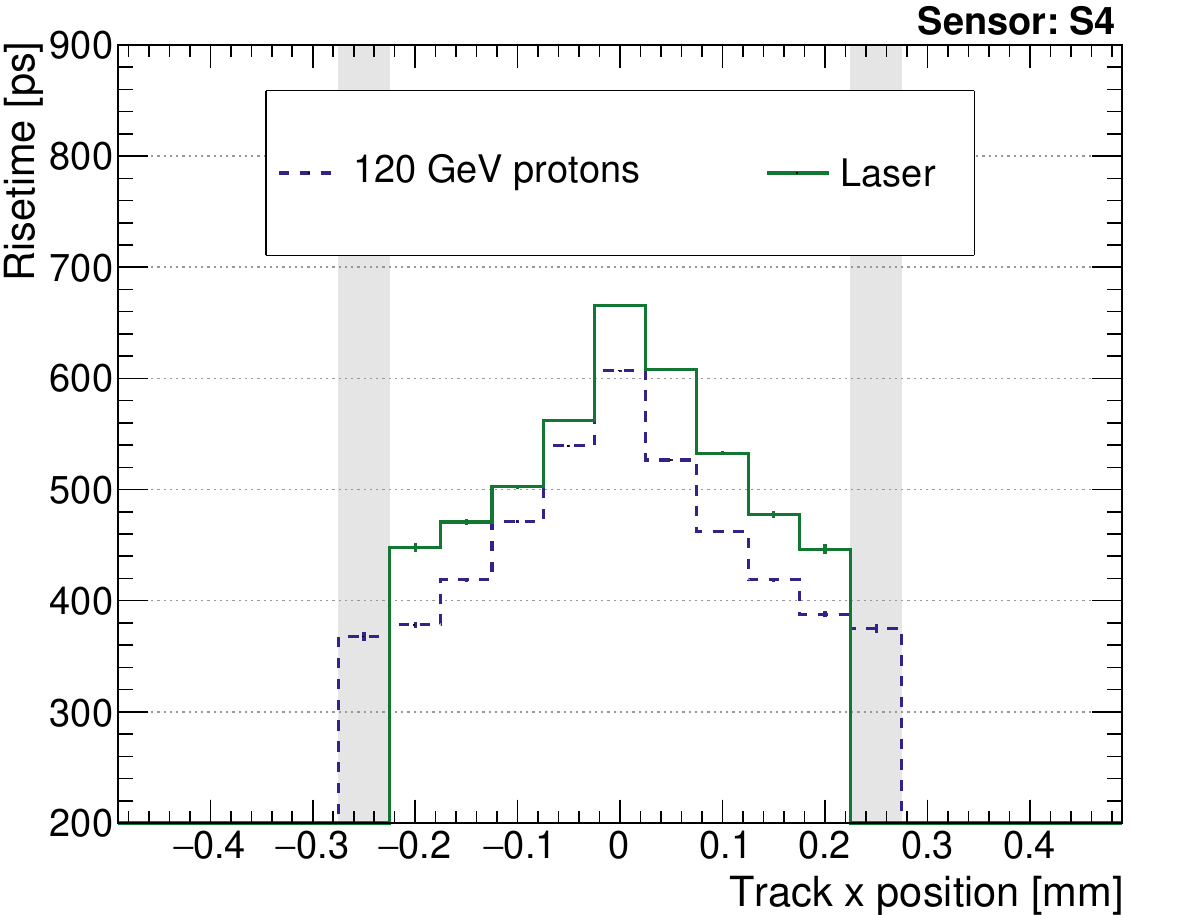}
        \caption{}
        \label{fig:risetime-comparison-W9}
    \end{subfigure}
    \caption{Risetime from leading channel as a function of $x$ position for the three strip sensors: S1, S2, and S4. Results from the 120 GeV proton beam have been reproduced with data from~\cite{Dutta:2024ugh}.}
    \label{fig:risetime-comparisons-all}
\end{figure}

The distributions of risetime from the leading channel as a function of track $x$ position are compared for the laser and MIP source in Figure~\ref{fig:risetime-comparisons-all}. After calibrating the laser intensity to match the signal amplitude of the MIP data, the risetime distributions are found to be comparable between the two sources. Consequently, for equal noise levels in both setups, the corresponding jitter values - being dependent on signal amplitude, risetime, and noise - are also expected to be comparable.

\begin{figure}[htp]
    \centering
    \begin{subfigure}{0.33\textwidth}
        \includegraphics[height=1.6in]{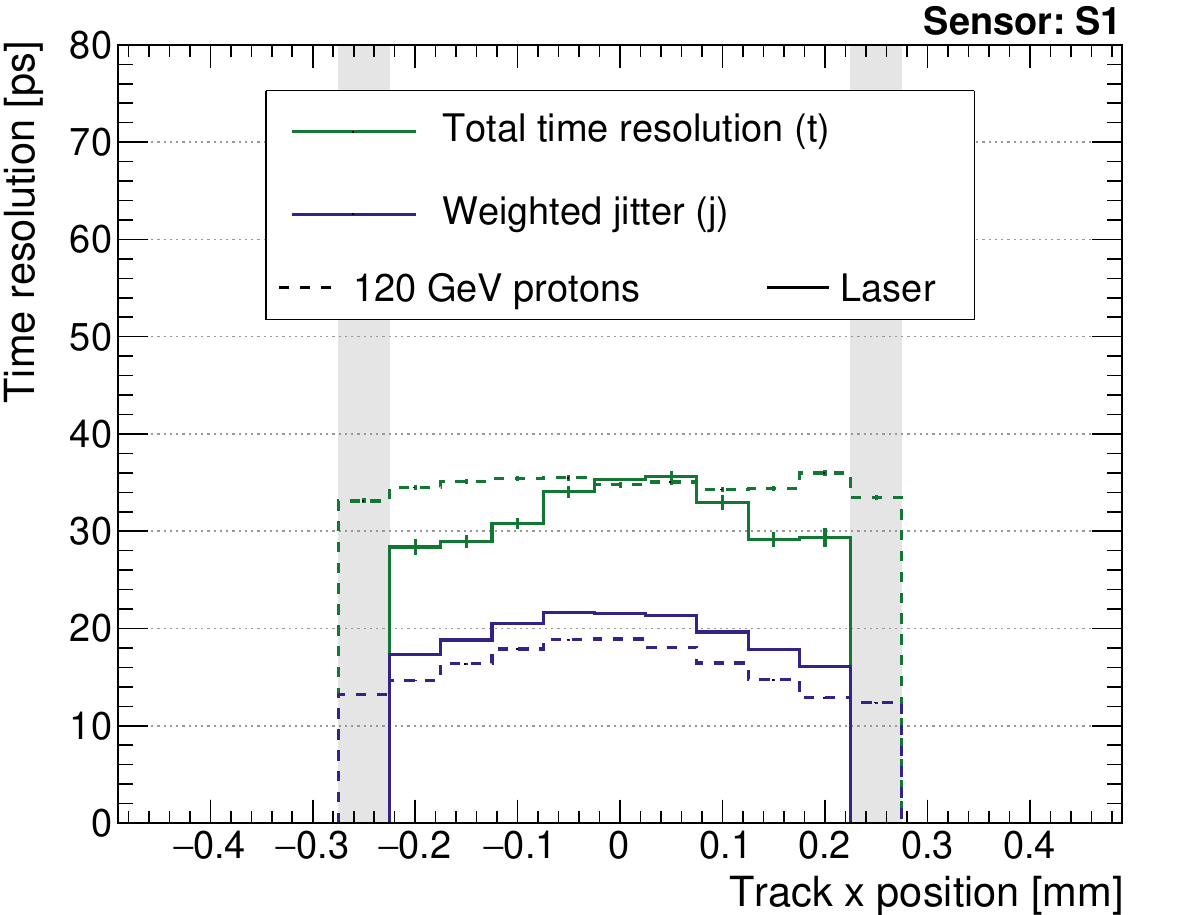}
        \caption{}
        \label{fig:tr-comparison-W2}
    \end{subfigure}
\hfill
    \begin{subfigure}{0.33\textwidth}
        \includegraphics[height=1.6in]{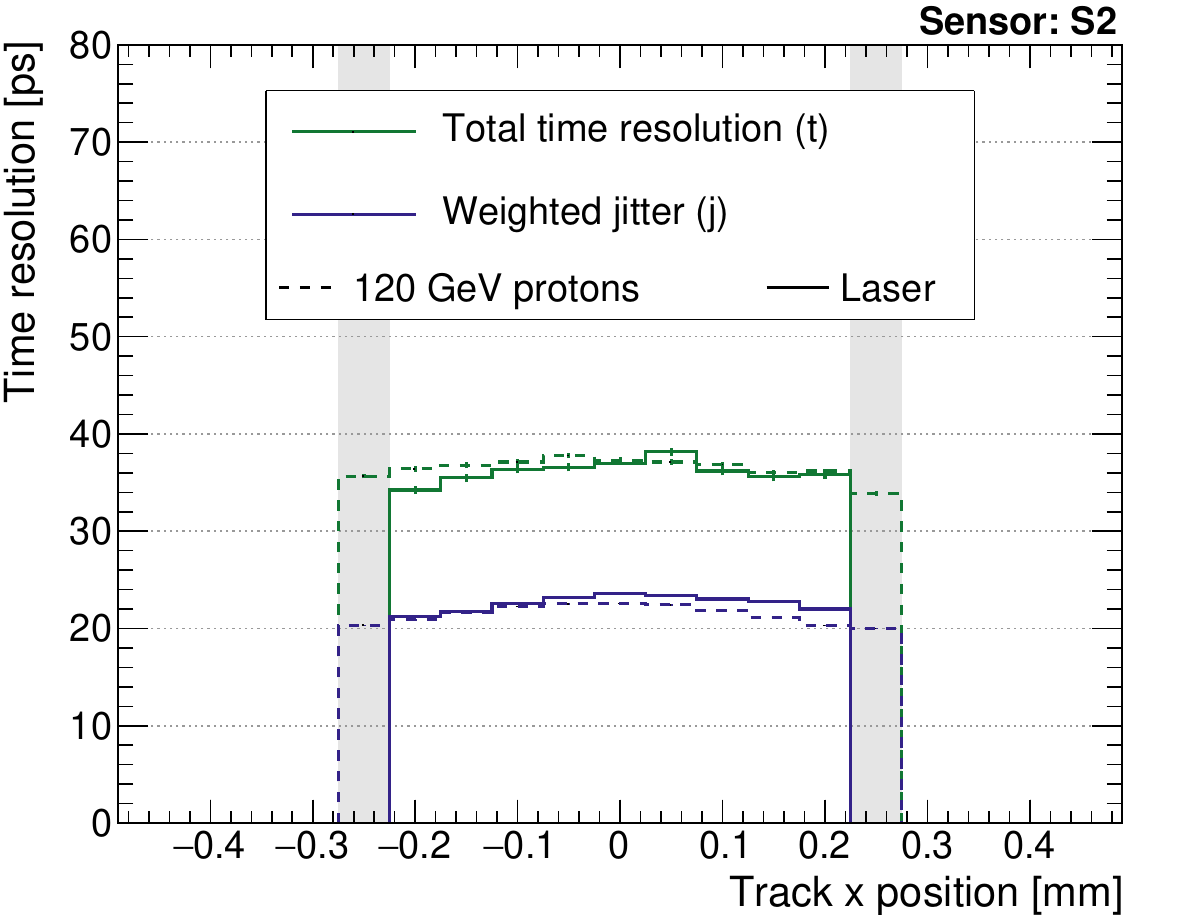}
        \caption{}
        \label{fig:tr-comparison-W4}
    \end{subfigure}
\hfill
    \begin{subfigure}{0.32\textwidth}
        \includegraphics[height=1.6in]{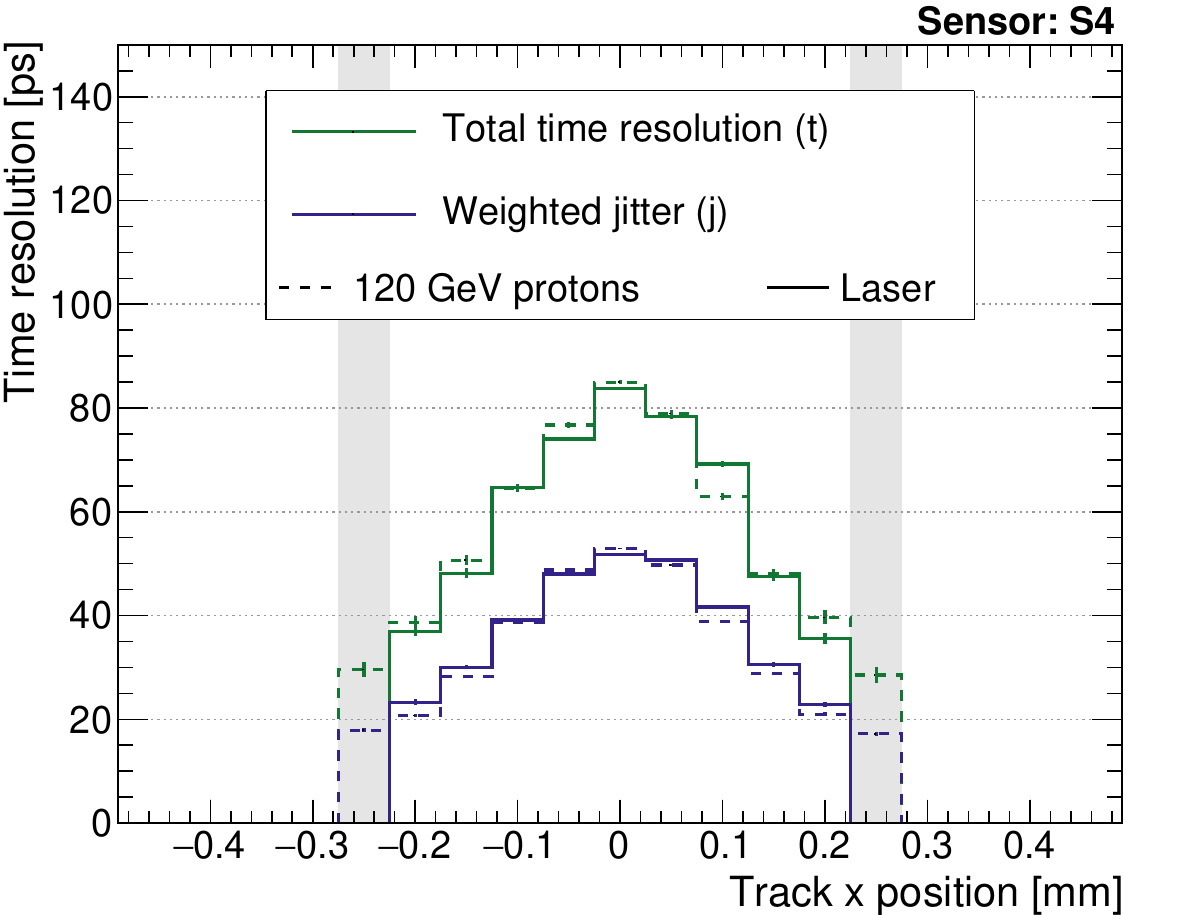}
        \caption{}
        \label{fig:tr-comparison-W9}
    \end{subfigure}
    \caption{Time resolution as a function of $x$ position for sensors S1, S2, and S4. The weighted jitter results from the laser source have been normalized to account for the noise differences between the two setups. Results from the 120 GeV proton beam have been reproduced with data from~\cite{Dutta:2024ugh}.}
    \label{fig:tr-comparisons-all}
\end{figure}

Timing measurements from both sources across all sensors are shown in Figure~\ref{fig:tr-comparisons-all} as a function of $x$ position, where the total time resolution is shown in green curves. Owing to different charge deposition mechanisms between MIP and infrared photons, the weighted jitter (plotted in the purple curves in Figure~\ref{fig:tr-comparisons-all}) is a more meaningful quantity to compare instead of total time resolution. The weighted jitter (j) results obtained using the laser source were normalized to account for noise differences between setups using Equation~\ref{eq:weighted-jitter} (i.e. normalization factor = $\frac{N_{\text{120 GeV protons}}}{N_{\text{Laser}}}$), and are found to be in reasonable agreement between the two sources across all sensors.

\section{Understanding components of time resolution}\label{sec:scale-factors}

The timing resolution in LGADs can be defined using the following equation~\cite{Sadrozinski_2018}:
\begin{equation}
    \sigma_t^2=\sigma_{\text{Jitter}}^2+\sigma_{\text {Landau }}^2+\sigma_{\text {Timewalk }}^2+\sigma_{\text {TDC }}^2
\end{equation}



Here, $\sigma_\text{Landau}$ component arises from fluctuations in charge deposition along the incident particle trajectory. The $\sigma_\text{TDC}$ is associated with the finite binning of Time-to-Digital Converters (TDCs). 
The sampling rate of the Teledyne LeCroy WaveRunner 8208HD oscilloscope is 10 GSa/s, and as described in Section~\ref{sec:time-reconstruction}, the timing measurements are extracted after fitting the rising edge of the waveforms. The expected contribution of $\sigma_\text{TDC}$ can be inferred from Figure~\ref{fig:sampling-rate-study} to be approximately 14 ps. The $\sigma_\text{Timewalk}$ contribution arising from variations in signal amplitudes due to Landau fluctuations, is suppressed by the use of the CFD method~\cite{Giacomini_2021} in the offline analysis.

After subtracting the weighted jitter from the total time resolution in quadrature in Figure~\ref{fig:tr-comparisons-all}, a non-negligible residual remains, which we collectively denote as the “additional component”. For measurements with MIP sources, this component in sensors S1 and S2 is consistent with the expected Landau fluctuations in 50~$\mu$m thick LGAD sensors~\cite{Sadrozinski_2018}. In contrast, the presence of a similar residual in the laser-source results cannot be fully explained. The non-zero constant (vertical offset) observed in the fits to the laser-source data points in Figure~\ref{fig:midgap-amp-vs-time-res} further indicates contributions from non-jitter sources to the measured total time resolution. We suspect that the conventional jitter formula (Equation~\ref{eq:single-channel-jitter}) may not accurately describe the observed behavior, or that additional, non-jitter effects influence the measurements.

\begin{figure}[htp]
    \centering
    \begin{subfigure}{0.33\textwidth}
        \centering
        \includegraphics[height=1.6in]{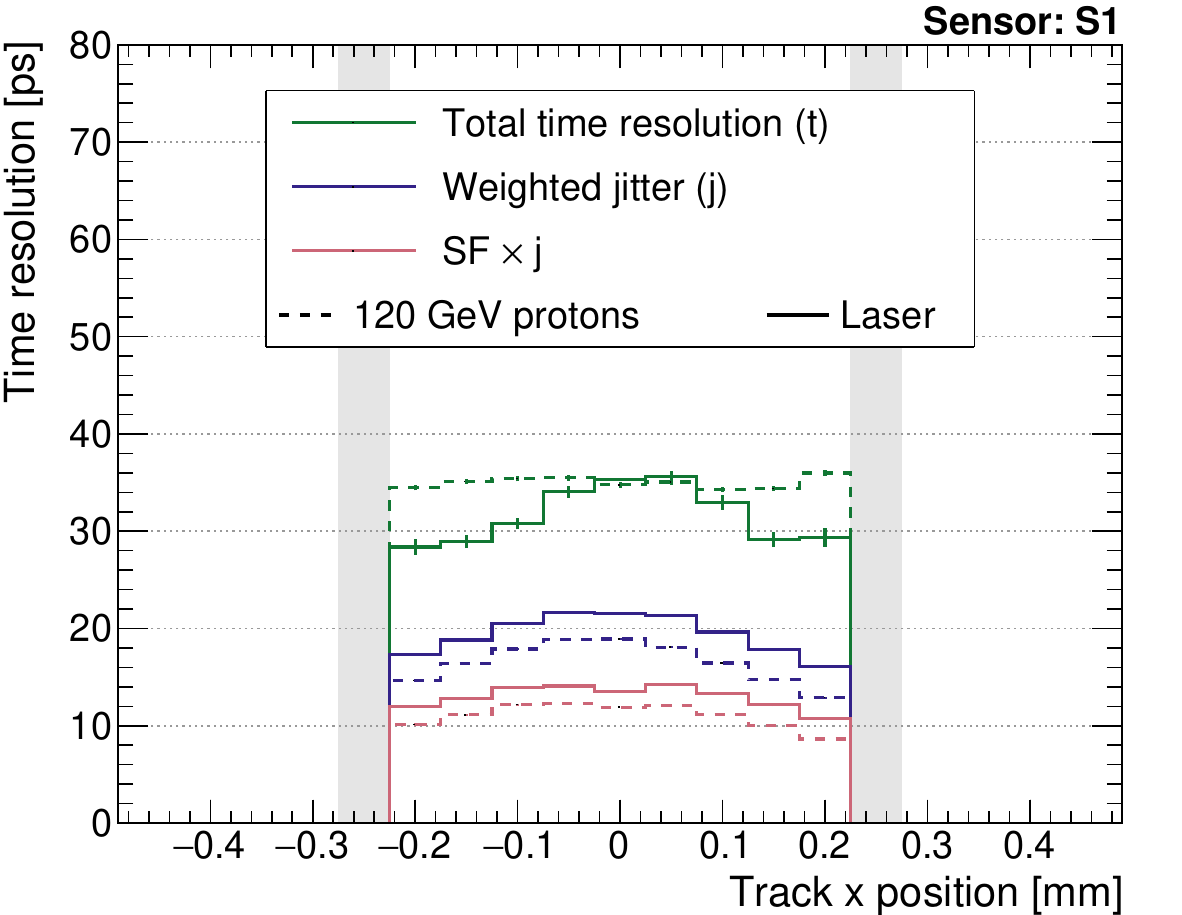}
        \caption{}
        \label{fig:scaled-tr-comparison-W2}
    \end{subfigure}
    \begin{subfigure}{0.33\textwidth}
        \centering
        \includegraphics[height=1.6in]{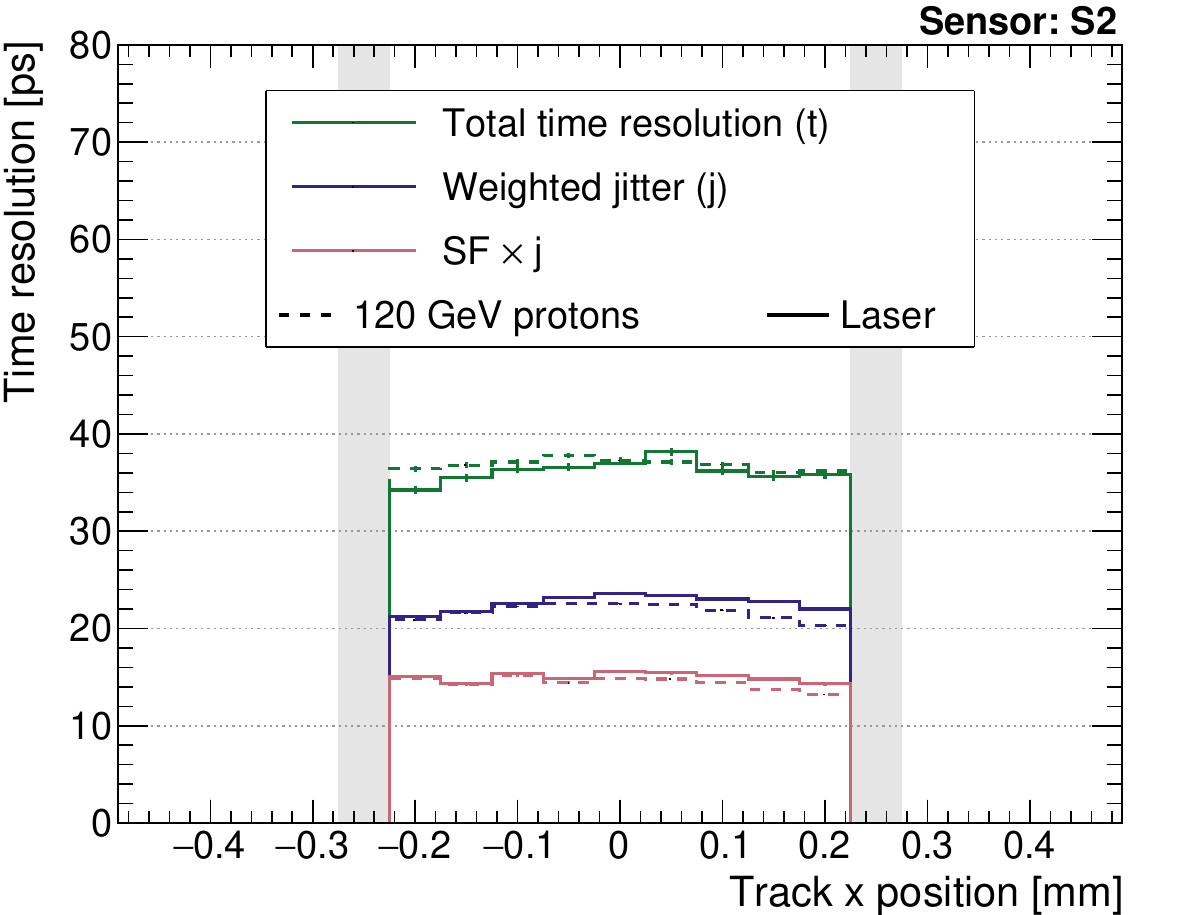}
        \caption{}
        \label{fig:scaled-tr-comparison-W4}
    \end{subfigure}
    \begin{subfigure}{0.32\textwidth}
        \centering
        \includegraphics[height=1.6in]{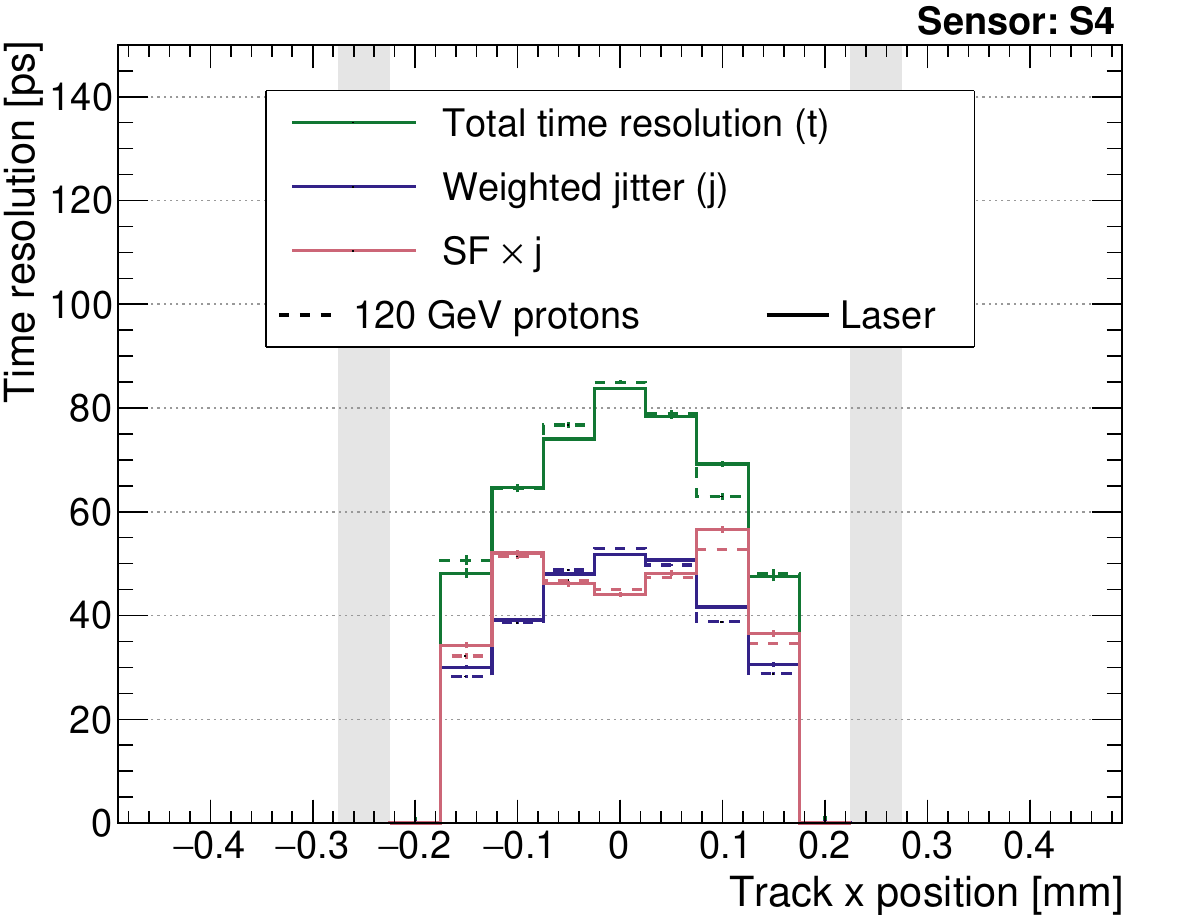}
        \caption{}
        \label{fig:scaled-tr-comparison-W9}
    \end{subfigure}
    \caption{Time resolution as a function of $x$-position for sensors S1, S2, and S4. The pink curves represent the weighted-jitter results after applying the scale factors (SF) obtained from simulation. Due to low statistics of the sub-leading channel's amplitude or risetime values for sensor S4, the histogram values at $x$ position = $\pm$0.2 mm is set to 0.}
    \label{fig:scaled-tr-comparisons-all}
\end{figure}

To account for possible inaccuracies in the jitter formula, scale factors derived from simulations (described in Section~\ref{sec:simulations}) were applied to the experimentally calculated weighted jitter results shown in Figure~\ref{fig:tr-comparisons-all}. The corresponding scaled results are presented as pink curves in Figure~\ref{fig:scaled-tr-comparisons-all}. 

A non-negligible additional component remains in all laser-source results, and in the MIP results for sensor S4, even after accounting for potential inaccuracies in the definition of single-channel jitter through this scaling procedure. This persistent discrepancy points to yet-unidentified mechanisms influencing time resolution across different sensor positions and warrants further investigation to fully understand their origin.

\section{Conclusions and outlook}\label{sec:discussion}
A test setup incorporating an infrared laser source was developed, enabling 2D scans of AC-LGAD sensors to assess their performance. This paper presented a technique for calibrating the laser intensity to mimic MIP-silicon interaction with data from a test beam campaign playing a role in calibrating the laser source and validating laser-based measurements. By accounting for setup-induced noise, we demonstrated that position resolution and weighted jitter results are consistent between the infrared laser and MIP proton sources across three AC-LGAD sensors. Thus, highlighting the reliability of laser setups to evaluate AC-LGAD sensor performance and their potential to expedite R\&D efforts by augmenting testbeam measurements.


The validated test setup now enables a broader range of semiconductor sensor characterization studies, including deeper investigations into the origins of the additional component in timing performance. Simulation studies together with experimental results indicate the need to understand non-jitter effects and refine the jitter estimation methodology. Additionally, several investigations outlined in~\cite{Bishop_2024} serve as examples of studies that can be conducted using the constructed setup. Future work involving a more diverse set of AC-LGAD sensors and a refined simulation framework would enhance our understanding of laser-based measurements and further validate our experimental program.

\section*{Acknowledgements}
This work was supported by funding from University of Illinois Chicago under Contract DE-FG02-94ER40865 with the U.S. Department of Energy (DOE). This document was prepared using the resources of the Fermi National Accelerator Laboratory (Fermilab), a U.S. Department of Energy, Office of Science, Office of High Energy Physics HEP User Facility. Fermilab is managed by FermiForward Discovery Group, LLC, acting under Contract No. 89243024CSC000002.

\bibliographystyle{elsarticle-num} 
\bibliography{references}{}

\end{document}